\documentclass[epj]{svjour} 
\usepackage{epsf,cite,epsfig,psfig,float,amssymb,stmaryrd,latexsym} 
\usepackage{graphics,psfrag}
\usepackage{graphicx} 
\newcommand{\be}{\begin{equation}} 
\newcommand{\ee}{\end{equation}} 
\newcommand{\beqa}{\begin{eqnarray}} 
\newcommand{\eeqa}{\end{eqnarray}}

\newcommand{\no}{\nonumber}

\begin{document} 
\title{Improved analysis of neutral pion electroproduction off deuterium in 
chiral perturbation theory 
\thanks{Work supported in part by   
Deutsche Forschungsgemeinschaft grants Me 864/16-2 and GL 87/34-1.}} 
 
\author{H. Krebs \inst{1} \thanks{Electronic address:~h.krebs@fz-juelich.de}  
\and 
V. Bernard \inst{2} \thanks{ 
Electronic address:~bernard@lpt6.u-strasbg.fr}  
\and 
Ulf-G. Mei{\ss}ner \inst{1}\inst{3} \thanks{ 
Electronic address:~meissner@itkp.uni-bonn.de}  
}                     
%
%
\institute{  
Helmholtz-Institut f\"ur Strahlen- und Kernphysik (Theorie), 
Universit\"at Bonn, Nu{\ss}allee 14-16, D-53115 Bonn, Germany 
\and 
Laboratoire de Physique Th\'eorique, 
Universit\'e Louis Pasteur, F-67084 Strasbourg Cedex 2, France 
\and 
Forschungszentrum J{\" u}lich, Institut f{\" u}r Kernphysik 
(Theorie),  D-52425  J{\" u}lich, Germany 
} 
\date{Received: date / Revised version: date} 
%
\abstract{ 
Near threshold neutral pion electroproduction on the deuteron is studied in 
the framework of heavy baryon chiral perturbation theory. 
We include the next--to--leading order corrections to the three-body 
contributions. We find an improved  description  of the total and differential 
cross section data measured at MAMI. We also obtain more precise values for the 
threshold S-wave multipoles. We discuss in detail the theoretical
uncertainties of the calculation.
} 
\maketitle 
\section{Introduction} 
\label{intro} 
Pion electroproduction off deuterium, $\gamma^\star (k) + d \to \pi^0 (q) + d$,
 allows to extract the elusive elementary
neutron amplitude. First data at small photon virtuality $k^2 = -0.1\,$GeV$^2$ and
small values of the pion excess energy $\Delta W$ ($\Delta W = 0.5, \ldots , 3.5\,$MeV)
have been taken at the Mainz Mikrotron MAMI-II \cite{Ewald}.
In the threshold region, this reaction can be analyzed in
the framework of chiral perturbation theory \cite{KBM1,KBM2}. In  Ref.~\cite{KBM2},
a (partial) next-to-leading order calculation was presented, which led to a
satisfactory description of the differential and total cross section data. That 
analysis can be improved in two respects. First, the so-called three-body corrections
(meson exchange currents) were only  considered at third order in that paper. Second,
a similar remark holds for the single scattering proton and neutron P-waves. In this
paper, we are concerned with the first issue, namely the next-to-leading (fourth) order
corrections to the three-body corrections. As will be shown, this introduces 
(in principle) no new
parameters and with that extension, the calculation is done with the same theoretical
precision as the fairly successful analysis of coherent neutral pion photoproduction off
the deuteron \cite{BBLMvK}. It also leads to an improved description of
the data as shown in this paper.
The second extension can not be done so straightforwardly. This is because so far only the
fourth order corrections to the photoproduction  P-wave multipoles are available
in the literature \cite{BKMp}. For the electroproduction case, similar calculations are
underway \cite{BKuM}, triggered in particular by the rather unexpected results for
neutral pion electroproduction off the proton at 
the photon virtuality $k^2 = -0.05\,$GeV$^2$
\cite{Merkel}. These data can not be described within chiral perturbation theory at
one--loop accuracy. This is in stunning contrast to the 
fairly good description of the data at  higher photon virtuality
from NIKHEF \cite{vandenBrink } and MAMI \cite{Distler} that was obtained
in \cite{BKMel}. Note
that remeasurements at the low photon virtuality seem to be more in 
line with expectations from chiral 
perturbation theory or sophisticated models \cite{MerkelCD}. As we will 
argue, the effect of such single scattering P-wave contributions can be 
effectively included in a refit of 
some four-nucleon-photon operators, which leads to a much improved description 
of the data. Given the precision of the data, a more elaborate treatment
appears inappropriate.
 
The manuscript is organized as follows. In Section~\ref{sec:form}, we briefly discuss the
pertinent formalism. We heavily borrow from Ref.~\cite{KBM2} and only discuss the new
fourth order three-body contributions in some detail. Our results are presented and
discussed in Section~\ref{sec:res} and we end with a short summary and outlook in 
Section~\ref{sec:sum}. Some technicalities are relegated to the appendices.

\section{Effective field theory description} 
\label{sec:form}
\subsection{General remarks} 
In Ref.~\cite{KBM2}, we developed the multipole formalism for 
near threshold pion electroproduction
off deuterium (see also \cite{Aren1,Aren2})
and calculated the pertinent transition matrix elements in chiral perturbation
theory. This is based on an effective chiral Lagrangian of pions and nucleons chirally
coupled to external sources like the photon,
\be
 {\mathcal L}_{\rm eff} =  {\mathcal L}_{\pi\pi} +  {\mathcal L}_{\pi N} + 
 {\mathcal L}_{N N}~,
\ee
where the first term subsumes the interactions between the Goldstone bosons, the
second the pion-nucleon interactions and the third term the short range part of
the two-nucleon interaction. All of these Lagragians are a series of terms with
increasing chiral dimension,
\beqa
 {\mathcal L}_{\pi\pi} &=&  {\mathcal L}_{\pi\pi}^{(2)} + {\mathcal L}_{\pi\pi}^{(4)} 
+ \ldots~, \\
{\mathcal L}_{\pi N} &=& {\mathcal L}_{\pi N}^{(1)} + {\mathcal L}_{\pi N}^{(2)} 
+ \ldots~, \\
 {\mathcal L}_{NN} &=&  {\mathcal L}_{NN}^{(0)} +  {\mathcal L}_{NN}^{(2)} 
 + \ldots~,
\eeqa 
where the ellipsis stand for terms of yet higher order. The explicit expressions for the
various terms are well documented in the literature, see e.g. \cite{BKMrev}.
The transition operators derived from this effective Lagrangian are then 
sandwiched between wave functions that were consistently
generated from chiral nuclear effective field theory. The latter we take from the
recent work of Ref.~\cite{EGMZ}. 
The various contributions can be organized in terms of a consistent power counting
in terms of a small parameter $q$, like e.g. a meson mass, energy or nucleon 
three-momentum (with respect to the typical hadronic scale, 
say the mass of the rho meson). For a generic matrix element one has,
\begin{equation} 
{\cal M}={q^\nu}{\cal F}(q/\mu), 
\end{equation} 
where $\mu$ is a renormalization scale and the function ${\mathcal F}$ is of order one. 
Furthermore, $\nu$ is a counting index, i.e. 
the chiral dimension of any Feynman graph, for the case of pion production
off nuclei, see e.g. \cite{BLvK}. In terms of this counting index, we include all terms
with $\nu = 0$ and $\nu = -1$ (see also \cite{KBM2} for details).

The transition
matrix elements are generated by two very different types of contributions,
\be
{\mathcal M} = {\mathcal M}^{\rm ss} + {\mathcal M}^{\rm tb}~,
\ee
where ``ss'' and ``tb'' denote the single-scattering and the three-body  contribution,
respectively. Here, single scattering means that the pion is emitted from the same
nucleon to which the photon couples with the other nucleon acting as a mere spectator.
Processes involving both nucleons are often called exchange currents, but we follow
here the notation due to Weinberg \cite{Weind}. At third order, all tb diagrams
involve graphs with one pion in flight, whereas at the order considered here, four-nucleon
contact terms with the photon absorption and pion emission from different nucleon legs
come in (as discussed in more detail below). These terms can be understood from 
integrating out heavier mesons, see e.g. \cite{Friar,EGME}. In such a picture,
such diagrams would thus correspond to heavy meson exchange currents. 

The ss terms are of course sensitive to
the elementary proton and neutron electroproduction multipoles, properly boosted to the
pion-deuteron center-of-mass frame (for details, see \cite{KBM2}). The proton
amplitude is fixed from the chiral perturbation theory study of Ref.~\cite{BKMel}.
As in our earlier
work, we perform two types of fits, where we have one respectively two undetermined 
parameters related to the elementary $n \pi^0$ amplitude. 
In the fits of type~1, we have one
free fifth order parameter, called $a_5^n$,  and we include the constraint from a 
leading order low-energy theorem to the fourth order counterterms 
(as explained in \cite{BKMel}). The numerical value for the LEC $a_3^n$ from
resonance saturation is -4.11~GeV$^{-4}$ (comprised of the contribution from the
$\Delta (1232)$ in static approximation and vector mesons).
 For the fits of type~2, we relax this constraint and thus
have two fit parameters, the  low-energy constants $a_3^n$ and  $a_4^n$. These two
fitting procedures give us a measure of the theoretical uncertainty at the order we
are considering. We also get another measure of the theoretical uncertainty by
considering chiral EFT at NNLO and varying the cut--off in the
Lippmann-Schwinger equation. This will lead to bands for the various
observables rather than to lines as e.g. in \cite{KBM2}. We will come back to 
this topic when discussing the results. We note already here that the
uncertainty related to the two fit procedures is sizeably bigger than the
one induced by the cut-off variation. 
  
In what follows, we will use as kinematical quantities the virtuality of the photon,
$k^2$, which is negative in electron scattering (in the literature one often uses 
the positive quantity $Q^2 = -k^2$), the photon polarization $\varepsilon$ and the
pion excess energy, $\Delta W$, that is the energy of the produced pion above threshold
in the $\pi-d$ center-of-mass system. For a more detailed discussion of the kinematics,
see e.g. \cite{Ewald}.

\subsection{Fourth-order three-body contributions} 
\label{sec:tb4}
\begin{figure*}[hbt] 
\vskip 0.3cm 
\epsfxsize=12cm 
\centerline{\epsffile{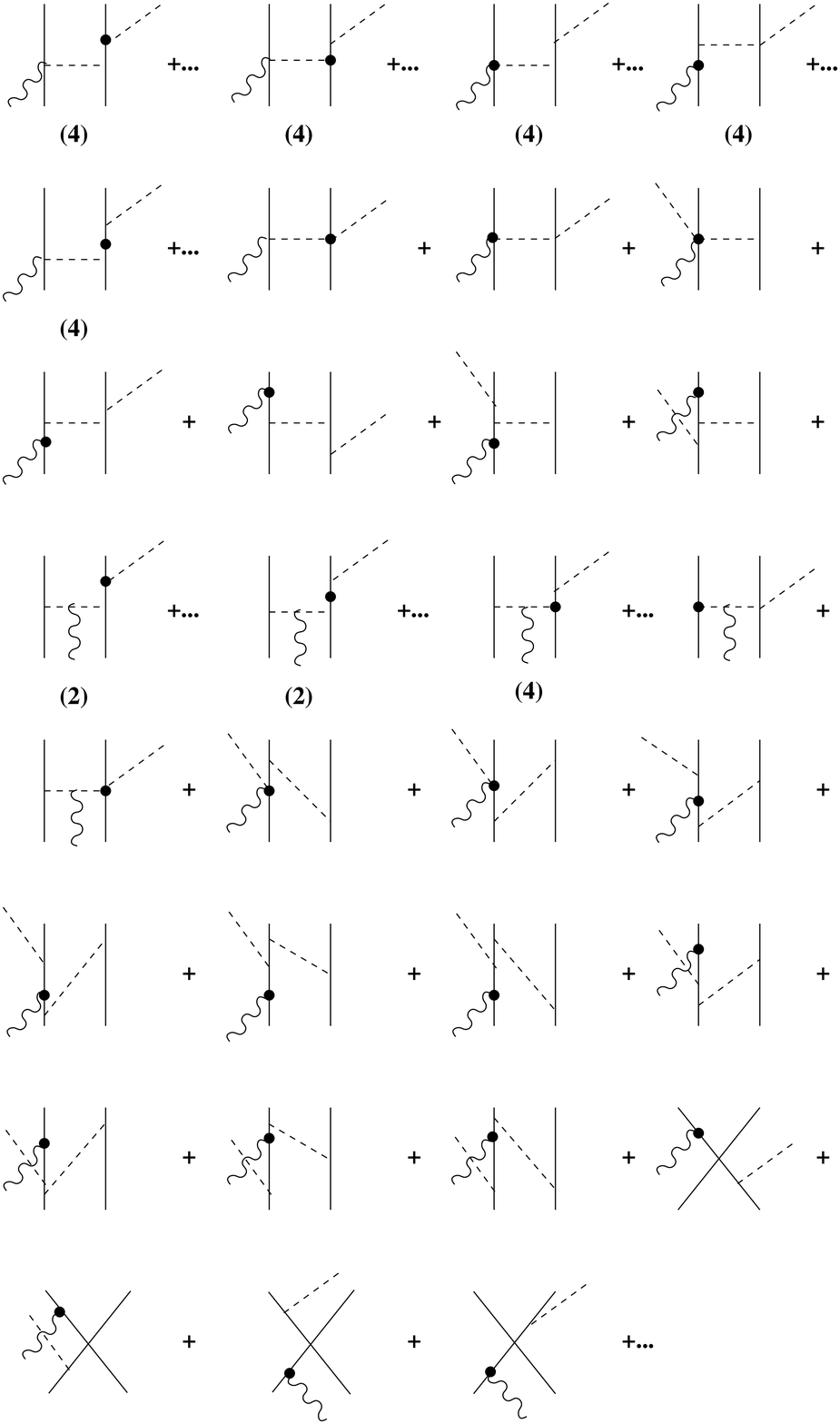}} 
\vskip 0.3cm 
\centerline{ 
\parbox{13cm}{\caption{Fourth order diagrams for the three-body corrections
(in Coulomb gauge).   Solid, dashed and wiggly lines  
denote nucleons, pions and photon, in order. The heavy dot stands for an insertion from 
the dimension two pion-nucleon Lagrangian. The number under certain graphs stands for the 
number of topological equivalent graphs as explained in the text.
The ellipsis at the end stands for Okubo-type corrections to the graphs involving
the four--nucleon contact interactions and also one-pion exchange.
\label{fig:diag4}} 
}} 
\end{figure*}

At third order, there are 8 tree graphs contributing to the three-body
corrections, see Fig.~4
of Ref.~\cite{KBM2} (using Coulomb gauge). At fourth order, there are altogether 59
non-vanishing diagrams, as shown in Fig.~\ref{fig:diag4}. These are  tree graphs
with exactly one insertion from the dimension two chiral pion--nucleon Lagrangian,
${\cal L}_{\pi N}^{(2)}$. 
More precisely, in that figure we have only shown the topologically inequivalent diagrams,
the numbers under certain graphs denote how many diagrams can be generated if one attaches
the pion emission vertex on the left or the right nucleon line above or below the pion 
exchange line. There is also a whole new class of short-distance diagrams including the 
leading (momentum-independent) four-nucleon interactions (the last four diagrams in 
Fig.~\ref{fig:diag4} and the corresponding Okubo-corrections, i.e. the diagrams
with two close-by energy denominators, are not shown. For details on this point,
see \cite{KBMokubo}. A more general discussion one these re-orthonormalization
diagrams can be found e.g. in \cite{EG,EGM1}.). 
The corresponding four-nucleon LECs have already been determined
in the fits to nucleon-nucleon scattering data. 
We use here the values collected in table~\ref{tab:Cvals} for the corresponding
NNLO wave functions.
\renewcommand{\arraystretch}{1.3} 
\begin{table}[htb] 
\begin{center} 
\begin{tabular}{|c||c|c|} 
\hline 
LEC   &  $\Lambda =450$~MeV  &  $\Lambda =650$~MeV  \\ 
\hline 
$C_{1S0}$ [$10^{-4}\,$GeV$^{-2}$] & $-$0.151 & $-$0.149 \\  
$C_{3S1}$ [$10^{-4}\,$GeV$^{-2}$] & $-$0.168 & $-$0.130 \\  
\hline\end{tabular}  
\vspace{0.1cm}
\caption{Values of the four--nucleon  LECs for the NNLO wave functions 
of \cite{EGMZ} used here. $\Lambda$ is the  cut--off in the
Lippman-Schwinger equation (the spectral function cut--off is fixed 
at $\tilde \Lambda = 650\,$MeV)
\label{tab:Cvals}} 
\end{center} 
\end{table} 

\noindent
However, these values stem from a fit to the NN phase
shifts at low energies. A fit including pion production data
(which is from the kinematical point of view closer to the process
considered here)
can lead to an increased theoretical uncertainty in the
determination of these LECs. To make that point more  transparent,
we briefly discuss the determination of the leading $4N\pi$ LEC
related to the D--term in \cite{EGM34}.  From low-energy observables
one obtains for this coupling in dimensionless units, $c_D/(F_\pi^2 \Lambda_\chi^2)
= 1.8 \ldots 3.6$,  whereas its determination
from P--waves in the reaction $pp \to pp \pi^0$ gives somewhat lower (but
still consistent) values, $c_D = 1.4 \ldots 2.1$
\cite{HvKM}~\footnote{Note that we have a different convention for the sign
of the axial-vector coupling $g_A$. The relation between the coupling constant
$\delta$ used in that paper and $c_D$ is given by $\delta = -0.143 c_D$.}. 
Also, at fourth order we will
have additional P--wave LECs from the single nucleon amplitudes.
With these two observations in mind, we will also perform
fits were we leave $C_{1S0}$ and $C_{3S1}$ as free parameters.
This will be discussed in more detail when we present the results.
Furthermore, ${\cal L}_{\pi N}^{(2)}$
contains some terms with fixed couplings (due to the constraints of Lorentz invariance)
and other terms with finite and scale-independent low-energy constants, denoted
$c_i$.  These LECs can e.g. be determined from the analysis of  low-energy
elastic pion--nucleon scattering data utilizing chiral perturbation theory.
From the hadronic couplings only the term  $\sim c_4$ contributes, we use the value 
$c_4 = 3.4\,$GeV$^{-1}$ \cite{BuM}.

\begin{figure*}[htbp] 
\vskip 1.3cm 
\epsfxsize=10cm 
\centerline{\epsffile{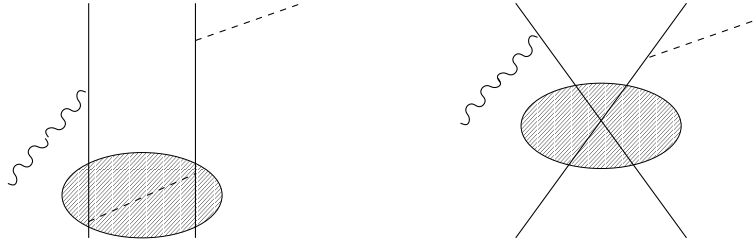}} 
\vskip 2.3cm 
\centerline{ 
\parbox{9cm}{\caption{
Classes of reducible diagrams as explained in the text. The shaded area
denotes the two-nucleon interaction, which includes either the leading
one-pion exchange (left) of the momentum--independent four-nucleon
interactions (right). Such types of diagrams are consistently not included
in the three-body interaction kernel.
\label{fig:irre}} 
}} 
\end{figure*}

\begin{figure*}[htbp] 
\vskip 3.cm 
\epsfysize=3cm 
\centerline{\epsffile{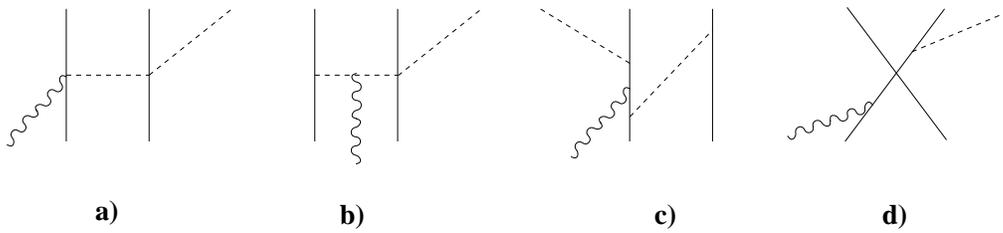}} 
\vskip 1.3cm 
\centerline{ 
\parbox{13cm}{\caption{
Classes of irreducible diagrams as explained in the text. 
a) represents a seagull, b) a pion-in-flight, c) a time-ordered and
d) a short-distance diagram, in order.
\label{fig:class}} 
}} 
\end{figure*}

We note that all the diagrams shown in  Fig.~\ref{fig:diag4} are irreducible,
where we use the following definition for irreducible diagrams:
A diagram is called irreducible, if it
contains no contributions from the two-nucleon potential (see e.g. the
related discussion in \cite{BMPK}).
This definitions becomes
obvious if one considers the Lippmann-Schwinger equation,
\beqa
\langle \phi_d \,\pi^0\,| T |  \phi_d \, \gamma^\star \rangle = 
\qquad\qquad\qquad\qquad && \no\\
\langle \phi_d \,\pi^0\,| (V_{\pi N}+ V_{\pi NN}) +  (V_{\pi N}+ V_{\pi
  NN})\qquad\qquad\qquad && \no\\
\times {1\over E-H_0- V_{NN}- V_{\pi N}- V_{\pi NN}- V_{\gamma N}- V_{\gamma
    NN} +i\epsilon} && \no\\
\times  (V_{\gamma N}+ V_{\gamma NN})   
|\phi_d \, \gamma^\star \rangle ~,&& 
\eeqa
and the deuteron wave function $\phi_d$ is obtained from the solution of the
Schr\"odinger equation,  
\begin{equation}
(H_0 +  V_{NN})~|\phi_d \rangle = E_d~|\phi_d \rangle~.
\end{equation}  
Here, $V_{\pi N}$, $V_{\gamma N}$, $\ldots\,$ are the pertinent pion-nucleon,
photon-nucleon, $\ldots\,$ transition potentials subject to the chiral
expansion as explained above.
This concept of reducibility is depicted in Fig.~\ref{fig:irre}, where two
typical reducible diagrams are shown, which include either the lowest order
one-pion-exchange
or the short-distance parts of the two-nucleon interaction. All diagrams
depicted in  Fig.~\ref{fig:diag4}
fall essentially in four classes, labeled a), b),c) and d). The
first two classes were already present at third order \cite{KBM2}, these are
the seagull-type and the pion-in-flight type graphs, respectively. Class c)
collects the so-called time-ordered graphs, were the photon couples to a
nucleon line and the pion is emitted from a nucleon while the exchanged pion
is in flight. All diagrams build from the four-nucleon contact terms are
collected in class d). Representative diagrams for these four classes are
shown in Fig.~\ref{fig:class}. In fact, a) falls into two subclasses,
depending on the value of the energy $q_0'$ of the exchanged pion. One either
has $q_0' = 0 + {\mathcal O}(1/m)$ or $q_0' = q_0 + {\mathcal O}(1/m)$, with
$q_0$ the energy of the produced pion. The diagrams are evaluated using
Fourier-transformation techniques, as briefly discussed in App.~\ref{app:fourier}. 
We also remark that all our coordinate space integrals are finite due to the
exponential fall-off of the chiral EFT wave functions. In contrast to what was
done in \cite{BBLMvK}, we thus do not need to introduce an additional cut-off
in the fourth order tb terms.

\section{Results and discussion} 
\label{sec:res}
In this section, we display the results for the multipoles, differential 
and total cross sections and the S--wave cross section $a_{0d}$ for the two fit 
strategies. We have performed calculations with the chiral EFT wave 
functions at NNLO \cite{EGMZ} for cut-offs 
in the range from 450 to 650 MeV (with the spectral function cut--off
$\tilde \Lambda$ fixed at 650~MeV, for details see \cite{EGMZ}).
Consequently,  for the observables we will obtain bands rather than single
line. This method of estimating the theoretical uncertainty is one of the
advantages of the chiral effective field theory approach employed here.
We note that at threshold we have performed the calculations in momentum and
in coordinate space, which serves as a good check on the nontrivial
numerical evaluation of the various integrals.

\subsection{Results with fixed four-nucleon LECs}
We first discuss the result obtained keeping the four--nucleon LECs
$C_{1S0}$ and $C_{3S1}$ fixed at the values given in section~\ref{sec:tb4}.
In table~\ref{tab:LECs} we collect the fitted LECs related to the
elementary neutron amplitude. As discussed before,  we have  performed two types
of fits. We see from the table that the wave function dependence of the LECs
is very weak, which points towards the conclusion that the pertinent matrix
elements are dominated by the contributions from the long--ranged pion
exchange. The resulting values are also not very different from the ones
obtained in \cite{KBM2}, which are also given in the table.
\renewcommand{\arraystretch}{1.3} 
\begin{table}[htb] 
\begin{center} 
\begin{tabular}{|c||c|c|c|} 
\hline 
LECs   &  $\Lambda =450$~MeV  &  $\Lambda =650$~MeV  & ${\cal O}(q^3)$ \\ 
\hline 
$a_3^n$ [GeV$^{-4}$] & \phantom{$-$}4.148 & \phantom{$-$}4.140& \phantom{$-$}4.767 \\  
$a_4^n$ [GeV$^{-4}$] & $-$7.085 & $-$6.730 & $-$5.644 \\ 
\hline  
$a_5^n$ [GeV$^{-5}$] & $-$38.71 & $-$37.05 & $-$27.51\\  
\hline\end{tabular}  
\vspace{0.1cm}
\caption{Values of the fitted  LECs for the NNLO wave functions 
from \protect\cite{EGMZ}.
The values for $a_{3,4}^n$ refer to the fits~2, whereas 
the corresponding $a_5^n$ belongs to the respective fits~1. 
For comparison, the LECs from \protect\cite{KBM2} are also shown.
\label{tab:LECs}} 
\end{center} 
\end{table} 

\noindent Before showing results for the pertinent observables, we discuss a
few general features of the new contributions. First, we observe that the
4N contact interactions contribute mostly to the P-waves, their S-wave contribution
is very small. Further, these P-wave contributions mostly feed into the
magnetic $M_{10}^1$ multipole and cancel to some extent the corresponding
contribution from the third order tb terms. 
Second, the other new diagrams also lead to more changes
in the P-wave multipoles, simply since at this order there are no free parameters.
We also note that the value of the constant $a_3^n$ in the fits~2 agrees nicely
in magnitude with the resonance saturation estimate, but comes out with opposite
sign.

\begin{figure*}[htbp] 
\epsfig{file=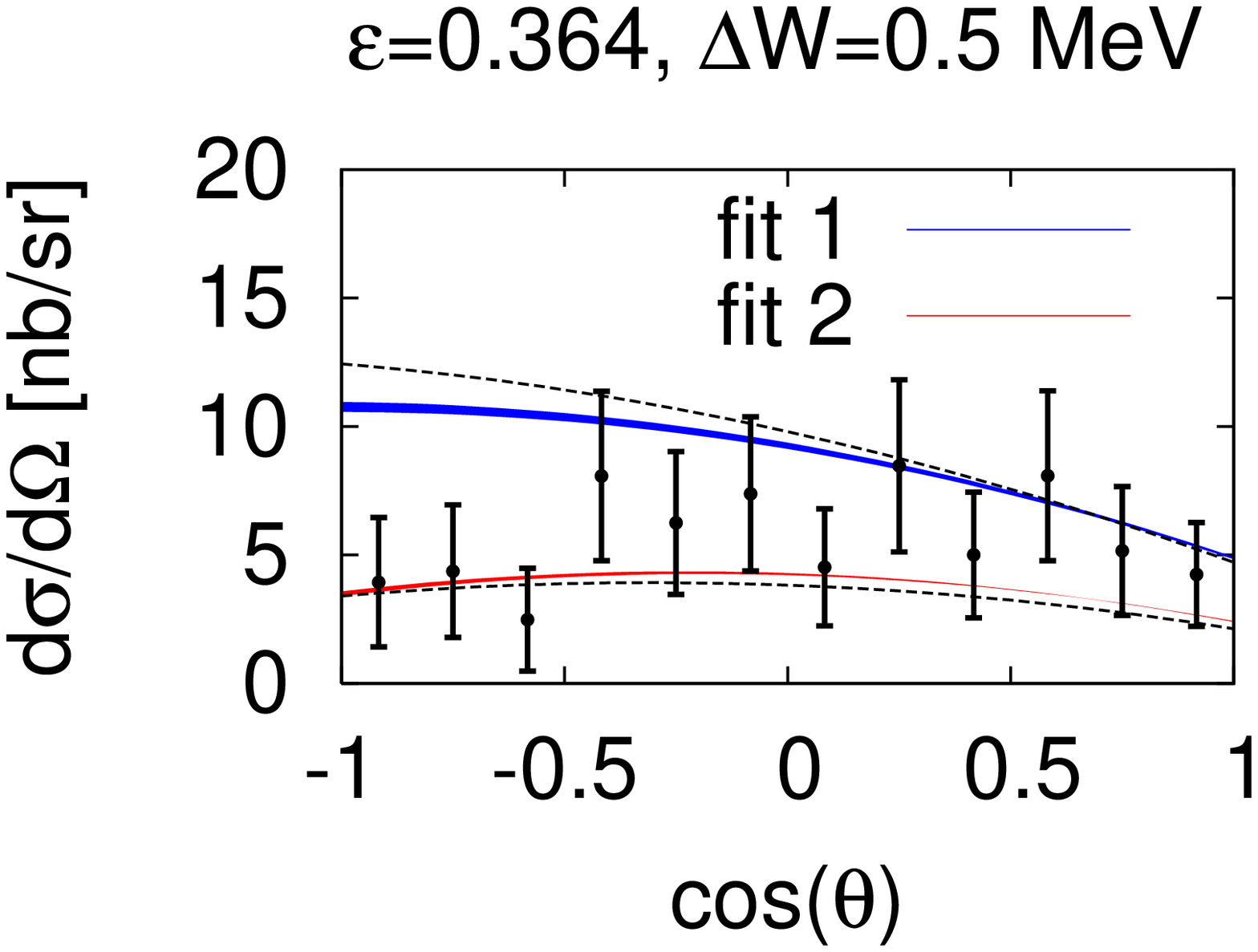,width=2.2in,angle=270} 
\hfill 
\epsfig{file=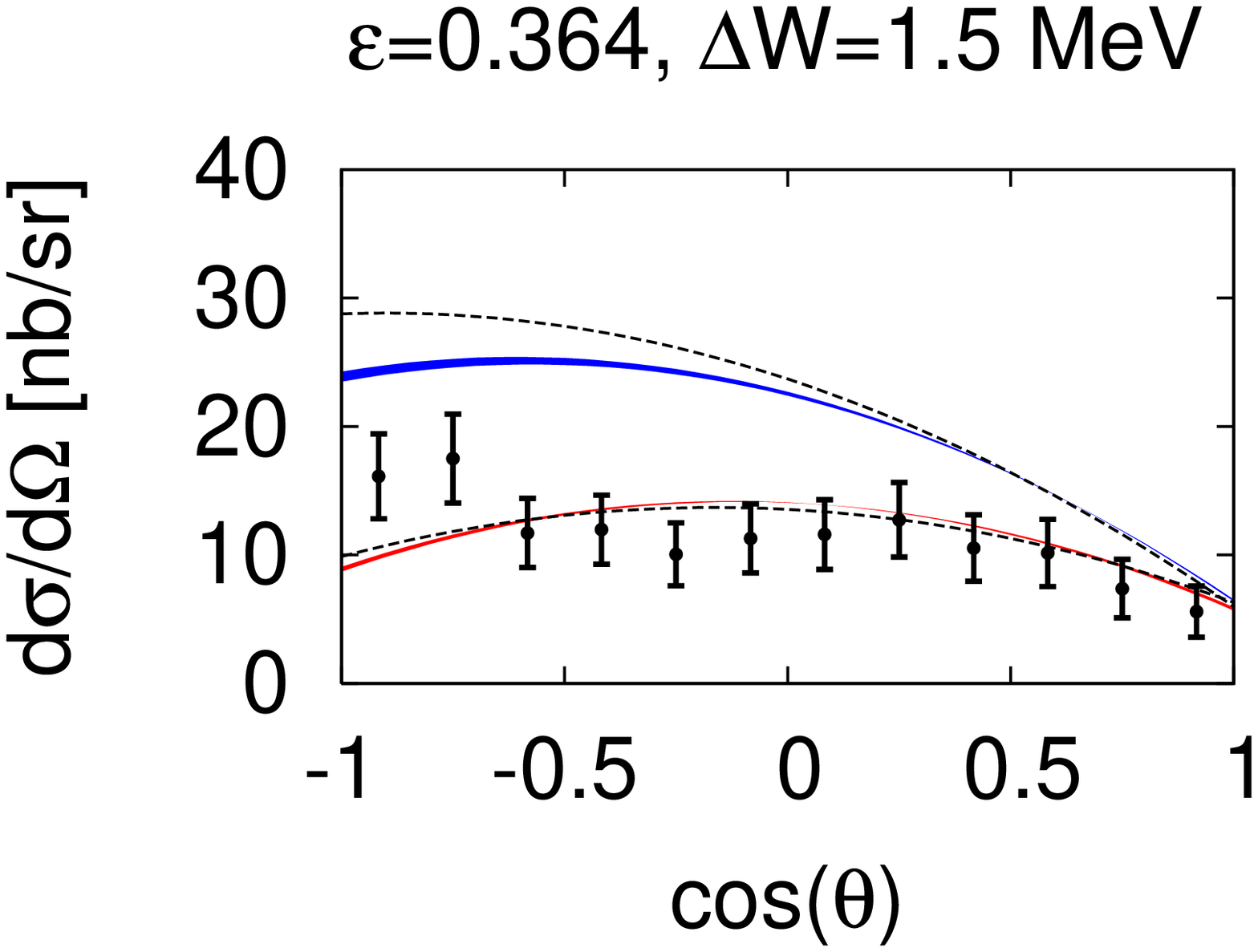,width=2.2in,angle=270} 
 
\vspace{0.5cm} 
 
\epsfig{file=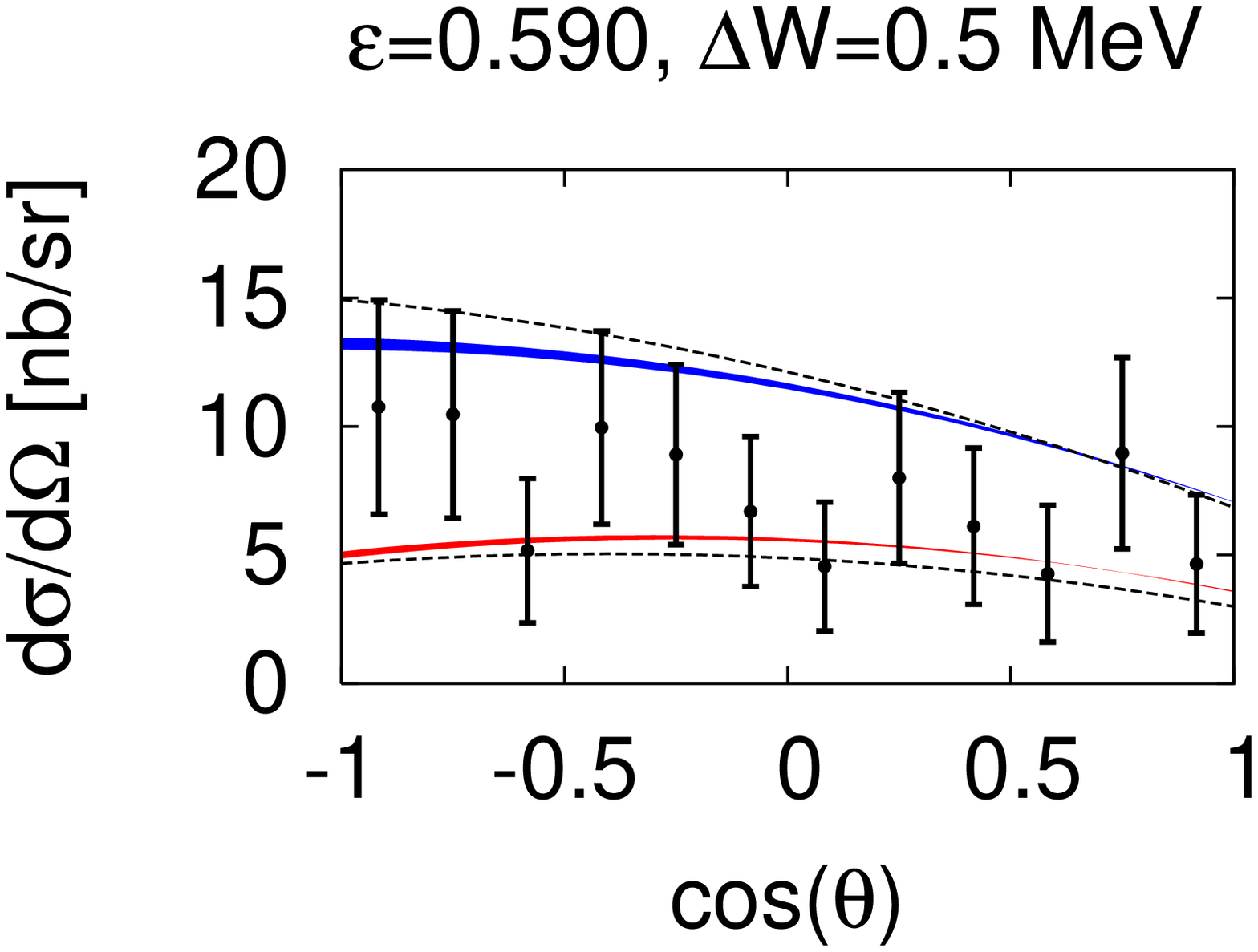,width=2.2in,angle=270} 
\hfill 
\epsfig{file=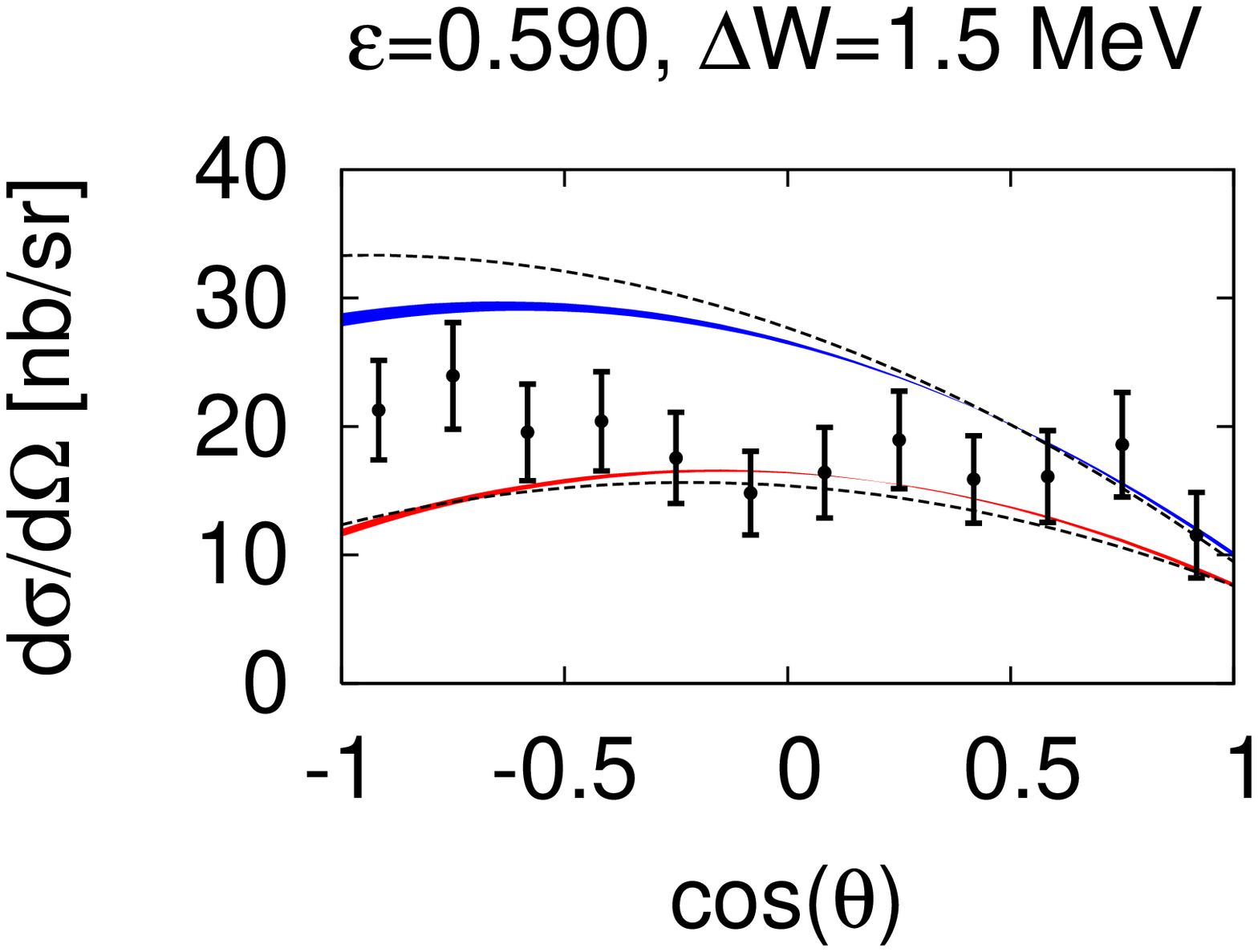,width=2.2in,angle=270} 
 
\vspace{0.5cm} 
 
\epsfig{file=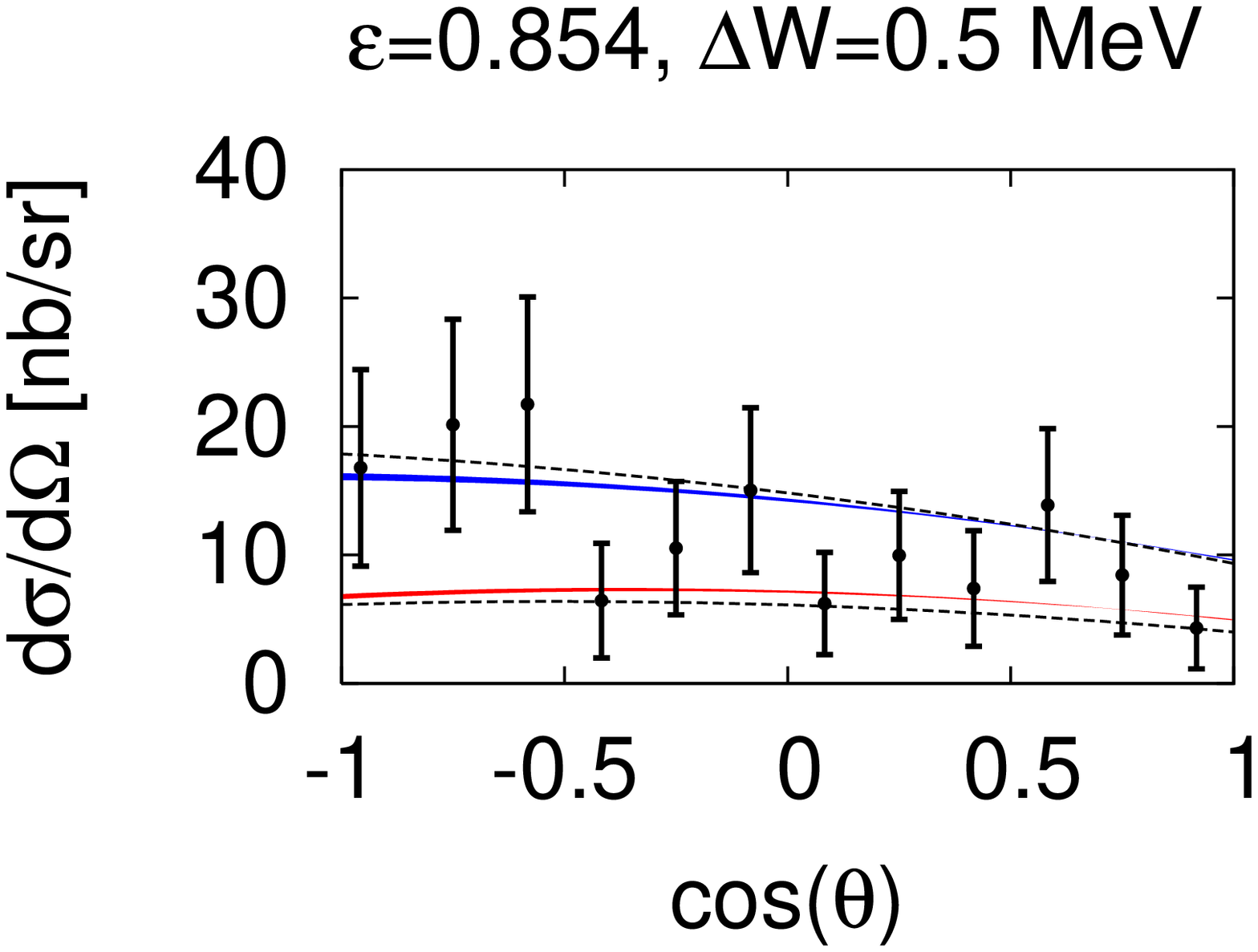,width=2.2in,angle=270} 
\hfill 
\epsfig{file=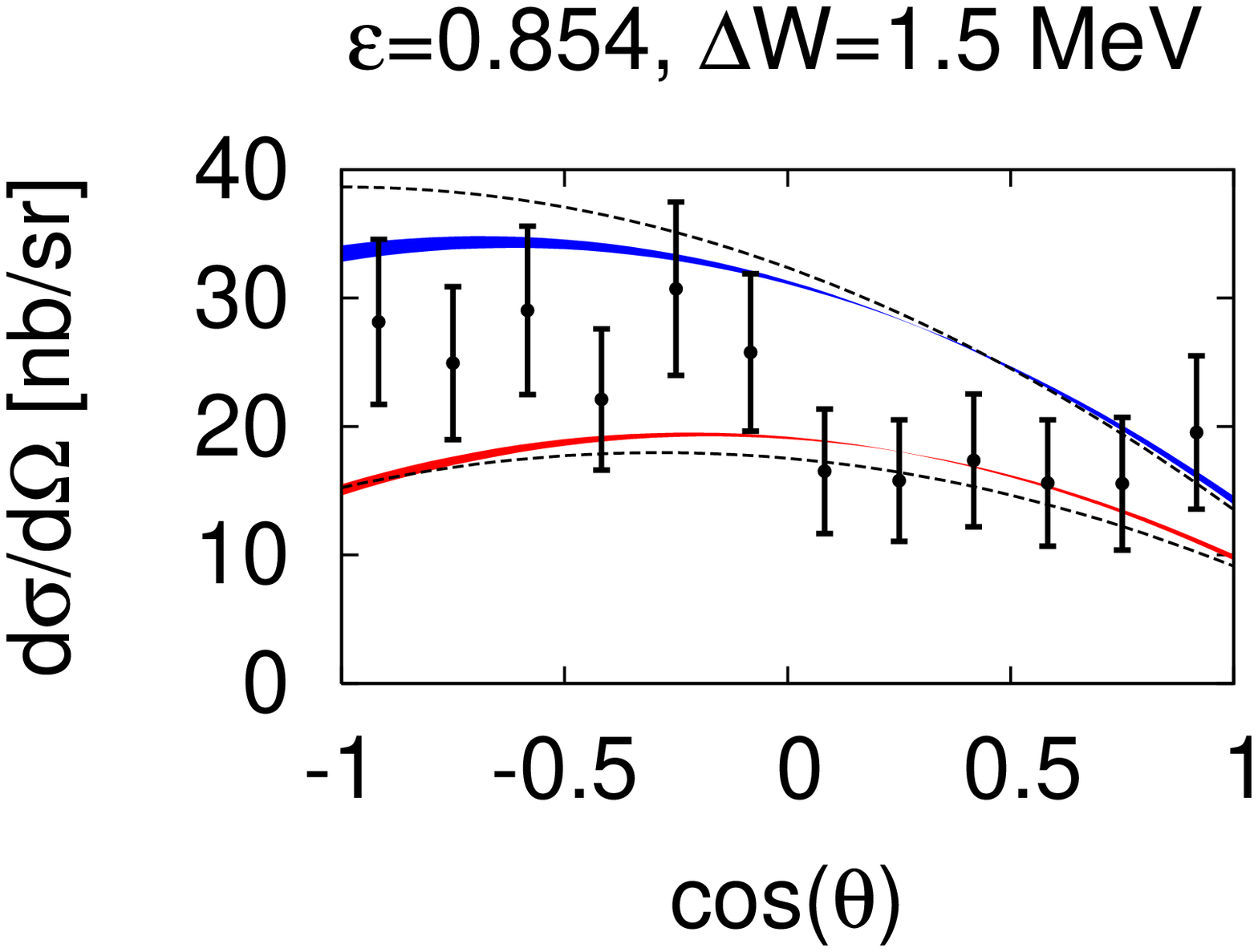,width=2.2in,angle=270} 
 
\vspace{1cm} 
 
\centerline{ 
\parbox{16cm}{\caption{Differential cross section at $\Delta W = 
    0.5\,$MeV (left column) and $\Delta W = 1.5\,$MeV (right 
    column) at three different values of the photon polarization 
    for the NNLO wave functions in comparison to the  
    MAMI data \protect\cite{Ewald}. Fits~1 and 2 are given
    by the blue (upper) and the red (lower) bands, respectively. 
    The bands are generated by varying $\Lambda$ from 450 to
    650 MeV. The dashed lines are the corresponding results 
    from \protect\cite{KBM2}.
    \label{fig:dXS0515}}} 
} 
\end{figure*}

\begin{figure*}[htbp] 
\epsfig{file=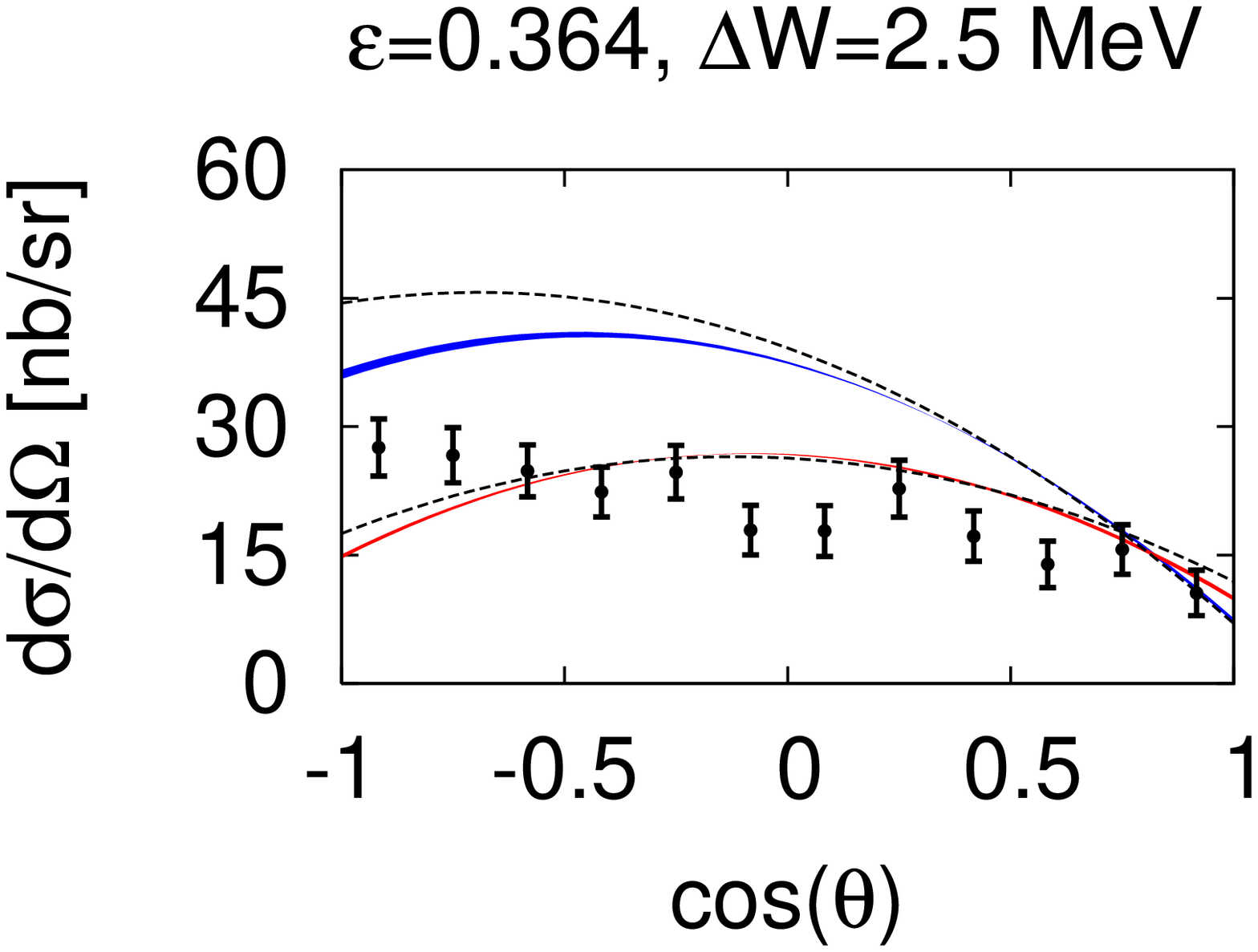,width=2.2in,angle=270} 
\hfill 
\epsfig{file=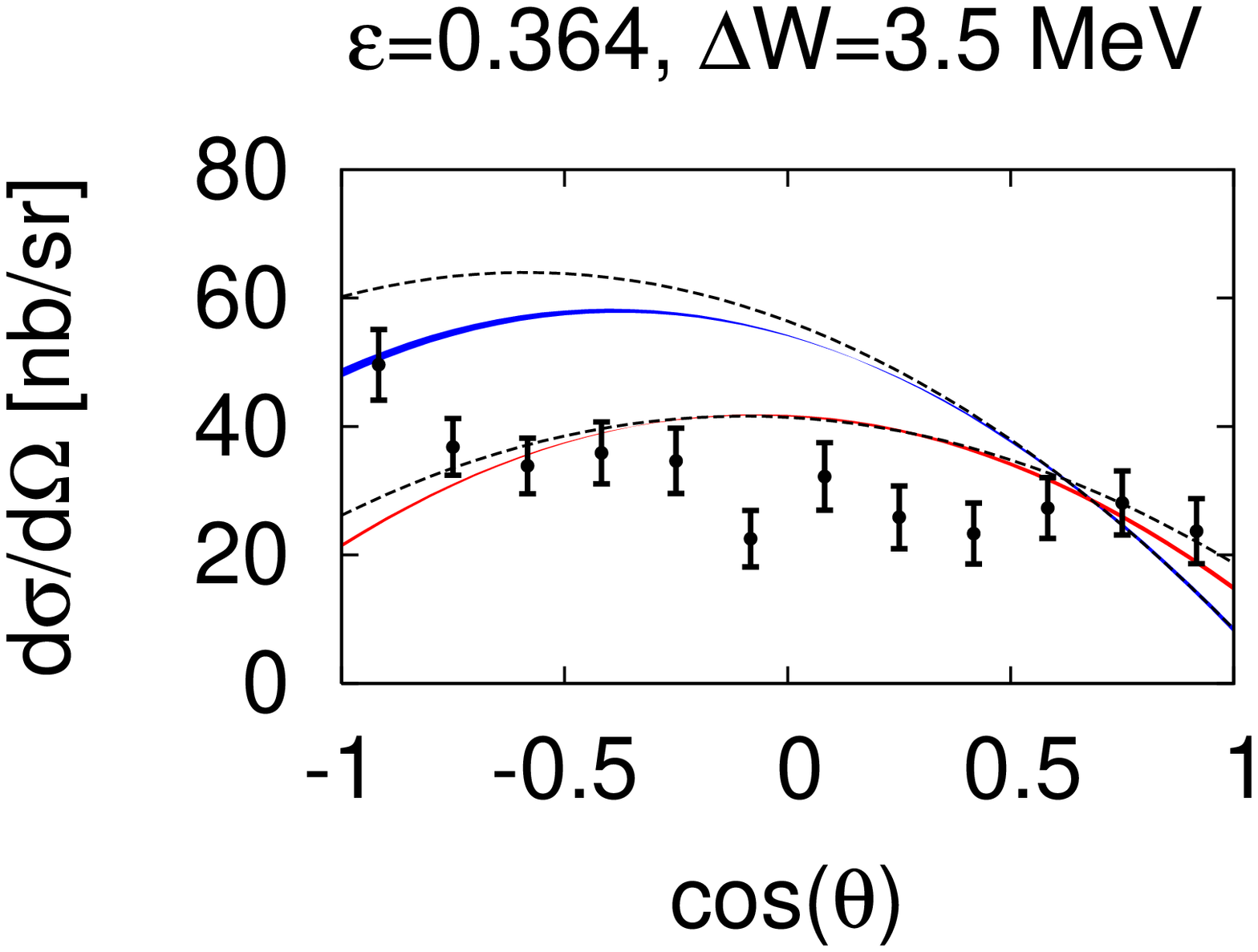,width=2.2in,angle=270} 
 
\vspace{0.5cm} 
 
\epsfig{file=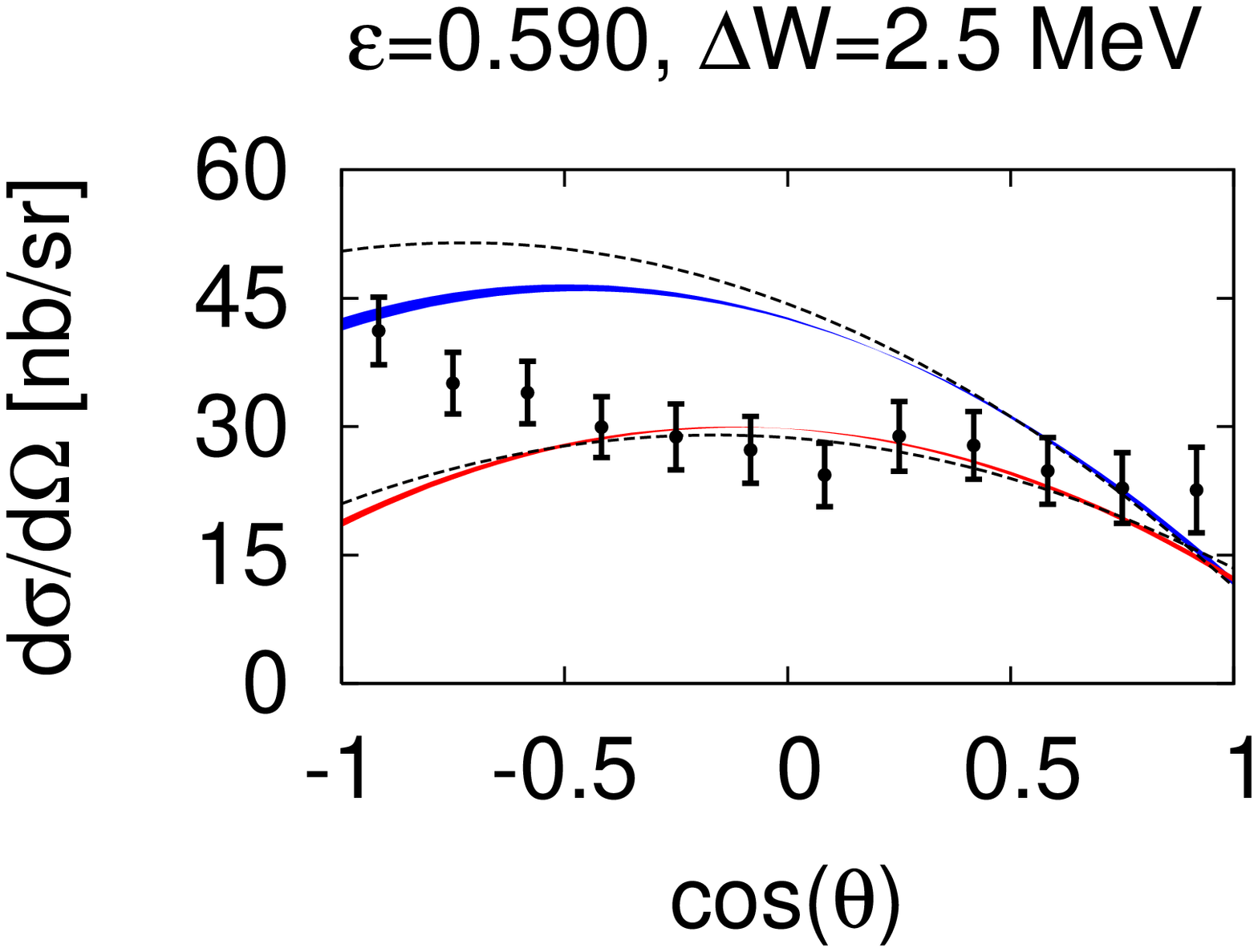,width=2.2in,angle=270} 
\hfill 
\epsfig{file=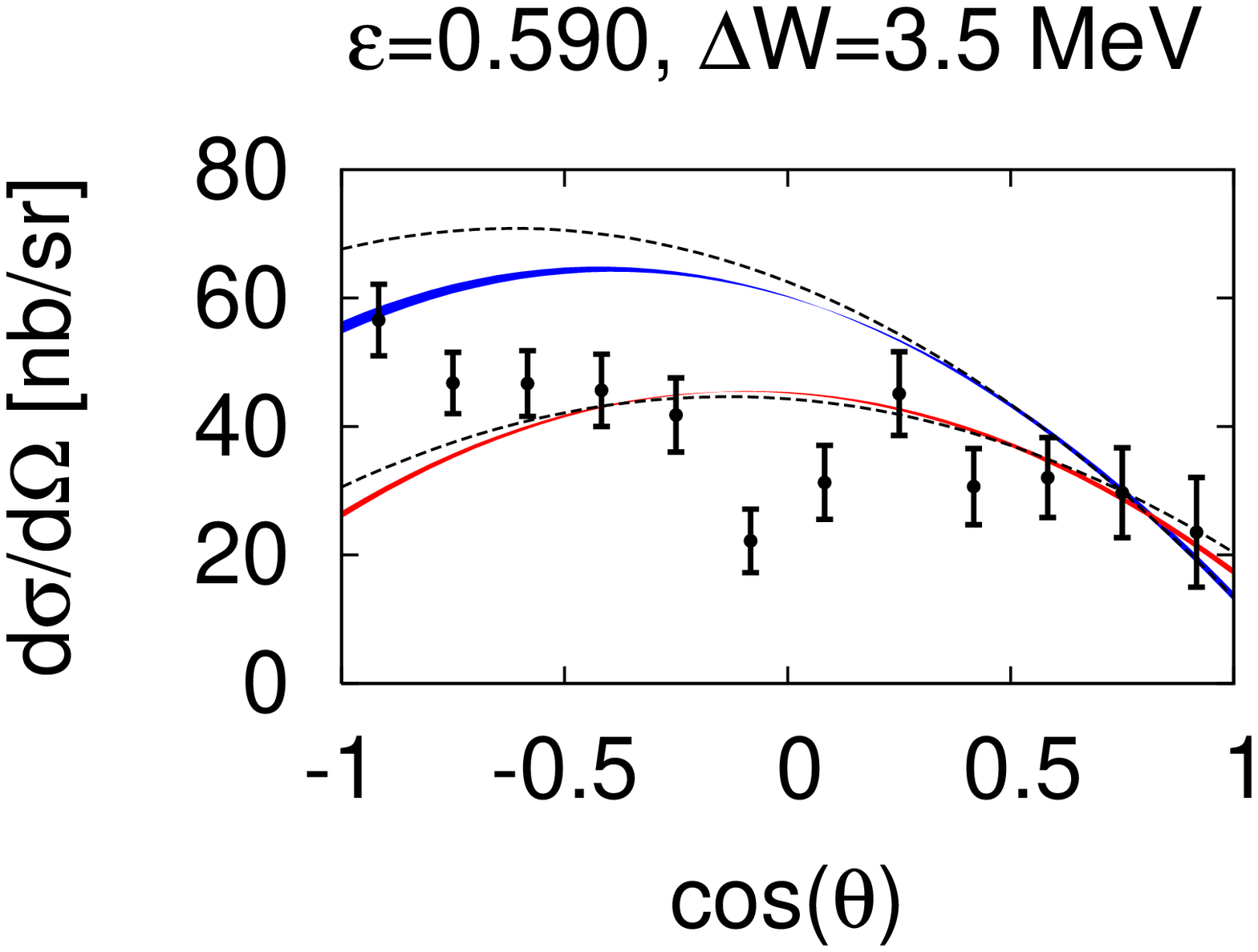,width=2.2in,angle=270} 
 
\vspace{0.5cm} 
 
\epsfig{file=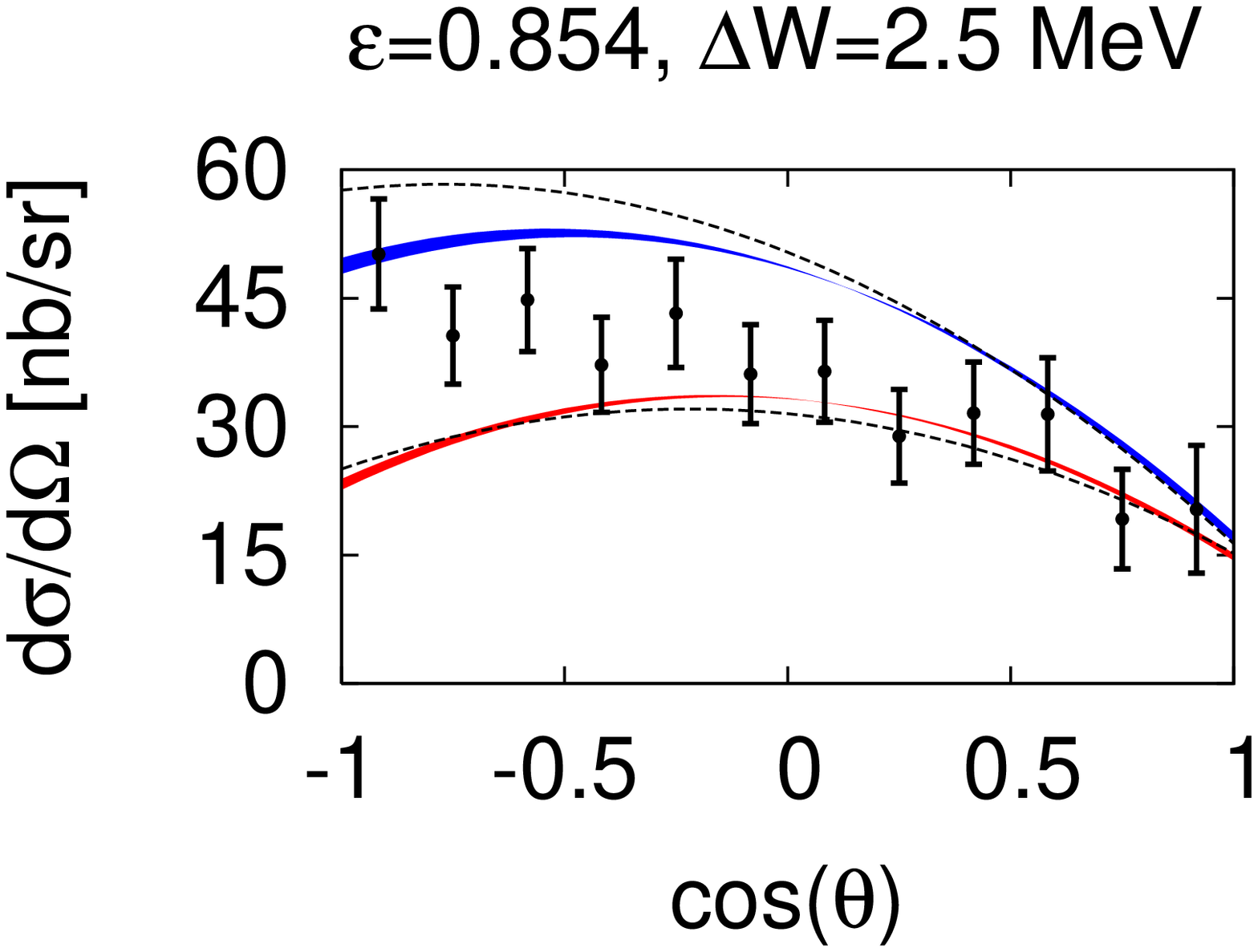,width=2.2in,angle=270} 
\hfill 
\epsfig{file=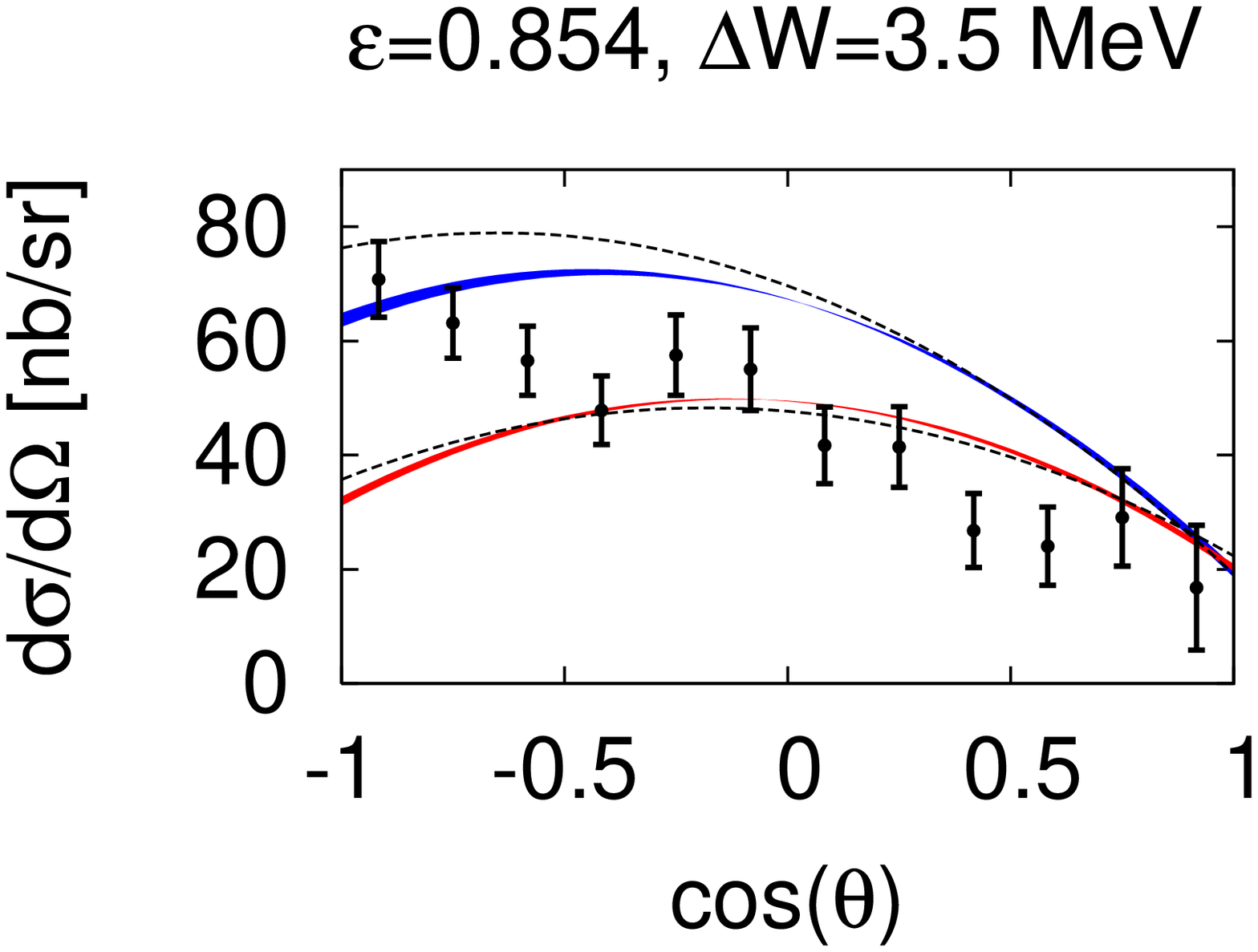,width=2.2in,angle=270} 
 
\vspace{1cm} 
 
\centerline{ 
\parbox{16cm}{\caption{Differential cross section at $\Delta W = 
    2.5\,$MeV (left column) and $\Delta W = 3.5\,$MeV (right 
    column) at three different values of the photon polarization 
    for the NNLO wave functions in comparison to the  
    MAMI data \protect\cite{Ewald}. Fits~1 and 2 are given
    by the blue (upper) and the red (lower) bands, respectively.
    The dashed lines are the corresponding results 
    from \protect\cite{KBM2}.
    \label{fig:dXS2535}}} 
} 
\end{figure*}

In Figs.~\ref{fig:dXS0515},\ref{fig:dXS2535} we show the differential cross sections 
for fits~1 and 2 employing the NNLO wave functions in comparison to the MAMI  
data \cite{Ewald}. These two bands (which are generated by utilizing the
NNLO wave functions with the cut-off $\Lambda = 450\,$MeV and 650~MeV) 
corresponding to the two fit procedures can be 
considered as a measure of the theoretical uncertainty at this order. We note
that the bands due to the wave function dependence are very thin, which shows
that the effective field theory calculation of these wave functions is of
sufficient accuracy.  The  uncertainty generated from the two fit procedures
is  comparable to  the experimental errors. 
We remark that the differences between the two fit procedures is somewhat
smaller as it
was the case when only the third order tb corrections were included \cite{KBM2},
see the dashed lines in  Figs.~\ref{fig:dXS0515},\ref{fig:dXS2535}.
Consequently, as it should be in a converging effective field theory, the theoretical
uncertainty has become smaller as compared to Ref.~\cite{KBM2}, although this
improvement is moderate.

\begin{figure*}[htpb] 
\begin{center}
   \vspace{0.5cm} 
    \includegraphics*[angle=270,width=0.45\textwidth]{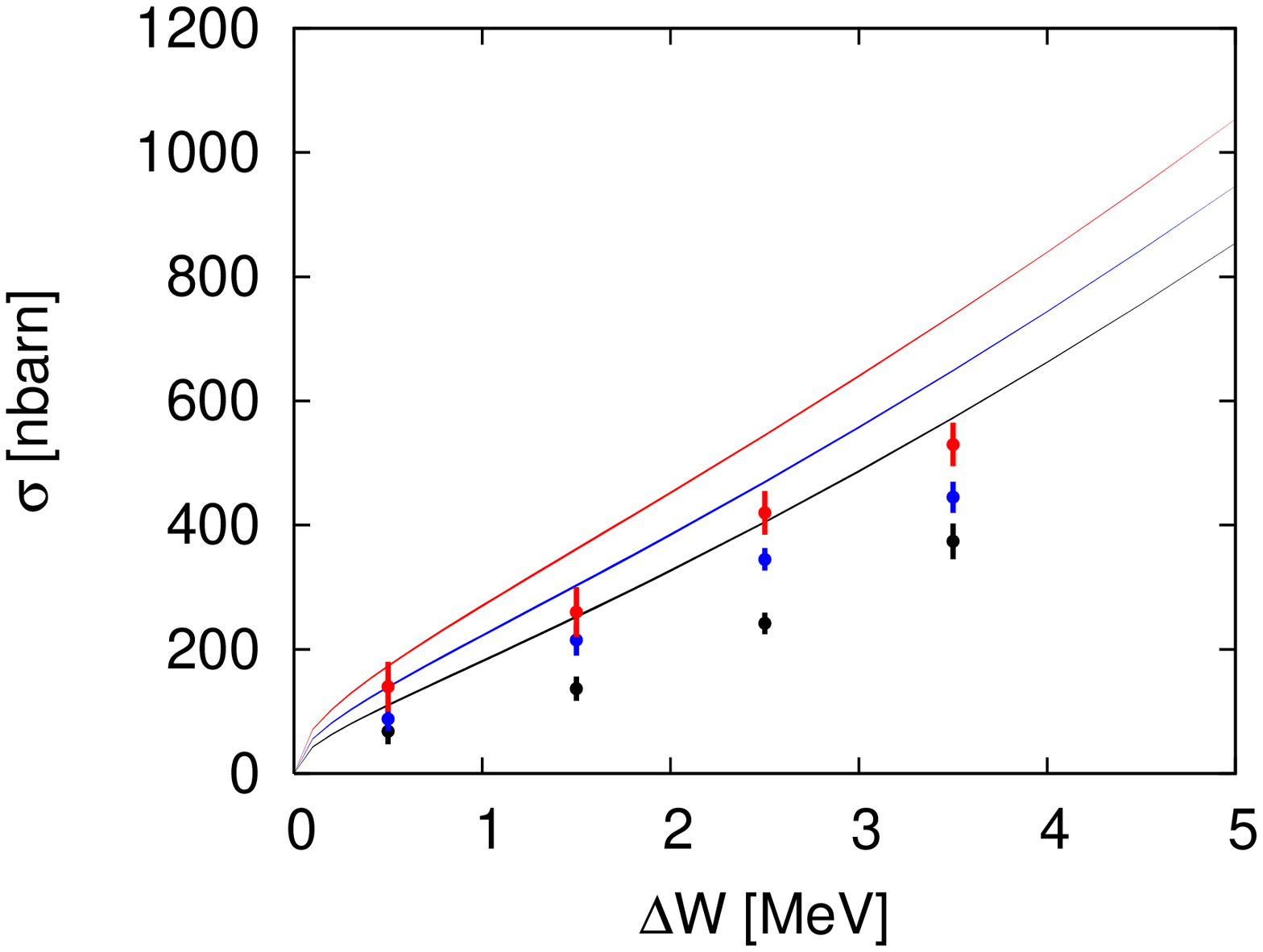}
    \includegraphics*[angle=270,width=0.45\textwidth]{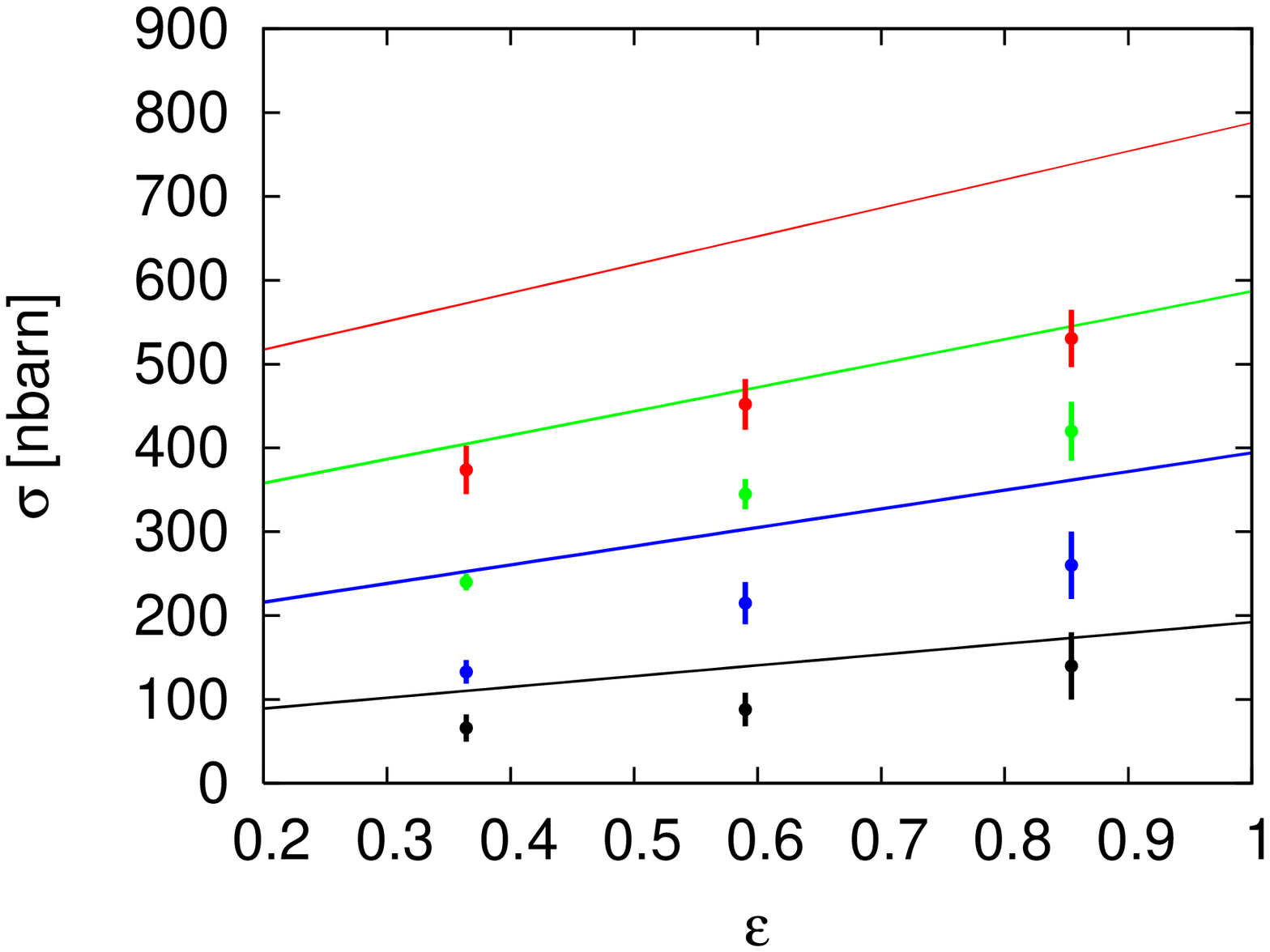}
   \vspace{0.7cm}
 \end{center} 
   \centerline{\parbox{12cm}{\caption{\label{fig:tot1} 
   Left panel: Total cross section as a function of $\Delta W$ for three 
   different values of the photon polarization in comparison 
   to the MAMI data \protect\cite{Ewald} for fit~1 and the 
   NNLO wave functions.  
   The upper/middle/lower band corresponds to the largest/medium/smallest 
   value of $\varepsilon$. 
   Right panel: Total cross section as a function of the photon polarization  
   $\varepsilon$ for four different values of the pion excess 
   energy $\Delta W$.
   The highest band corresponds to the largest value of $\Delta W$ and 
   correspondingly  for the other bands/values of $\Delta W$. 
  }}} 
\end{figure*}

\begin{figure*}[htbp] 
\begin{center}
   \vspace{0.5cm} 
    \includegraphics*[angle=270,width=0.45\textwidth]{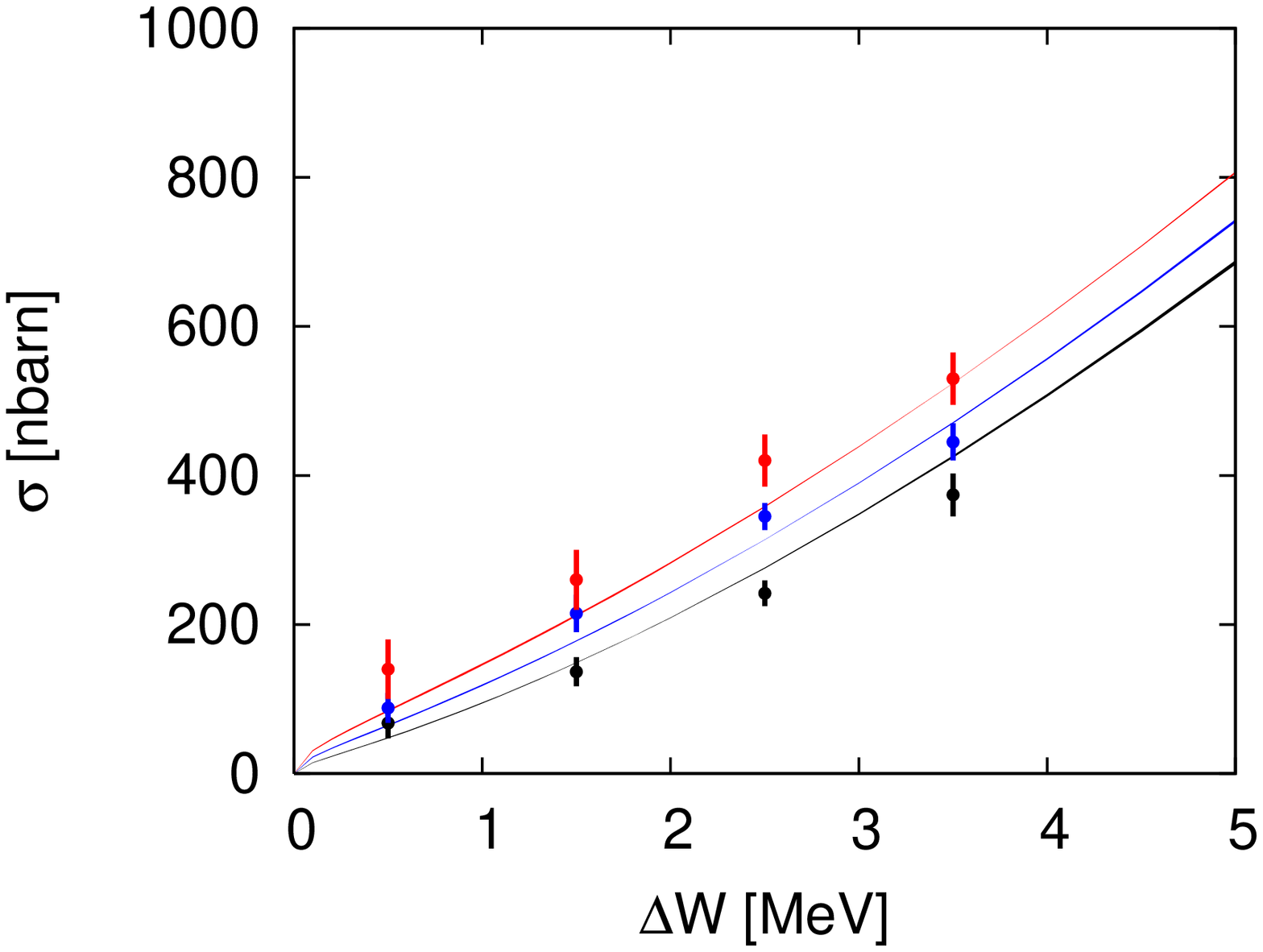}
    \includegraphics*[angle=270,width=0.45\textwidth]{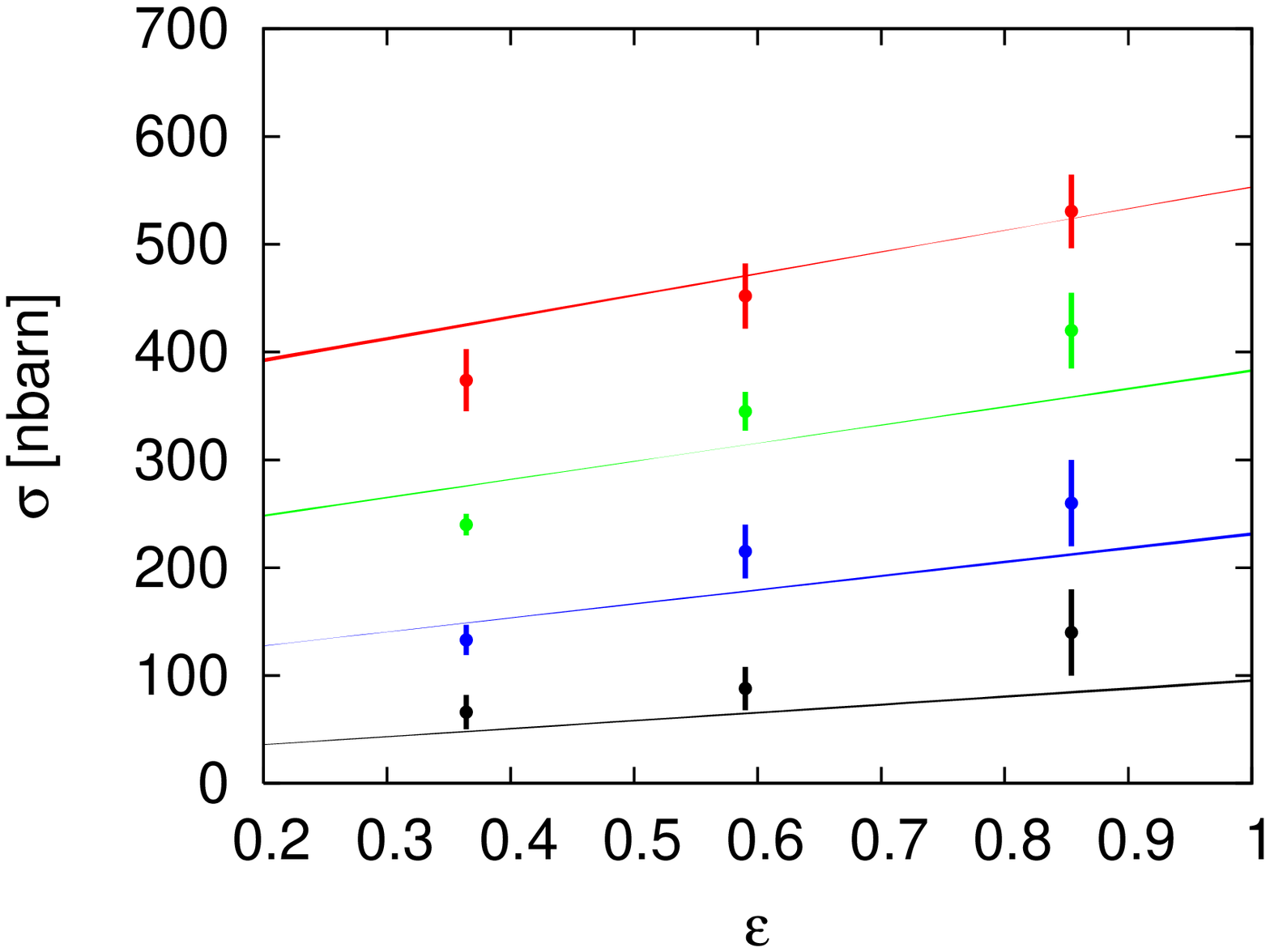}
   \vspace{0.7cm}
 \end{center} 
   \centerline{\parbox{12cm}{\caption{\label{fig:tot2} 
   Left panel: Total cross section as a function of $\Delta W$ for three 
   different values of the photon polarization in comparison 
   to the MAMI data \protect\cite{Ewald} for fit~2 and the 
   NNLO wave functions. 
   The upper/middle/lower band corresponds to the largest/medium/smallest 
   value of $\varepsilon$. 
   Right panel: Total cross section as a function of the photon polarization  
   $\varepsilon$ for four different values of the pion excess 
   energy $\Delta W$.
   The highest band corresponds to the largest value of $\Delta W$ and 
   correspondingly  for the other bands/values of $\Delta W$. 
  }}} 
\end{figure*}

\begin{figure*}[htbp] 
   \vspace{0.5cm} 
  \begin{center}
    \includegraphics*[angle=270,width=0.45\textwidth]{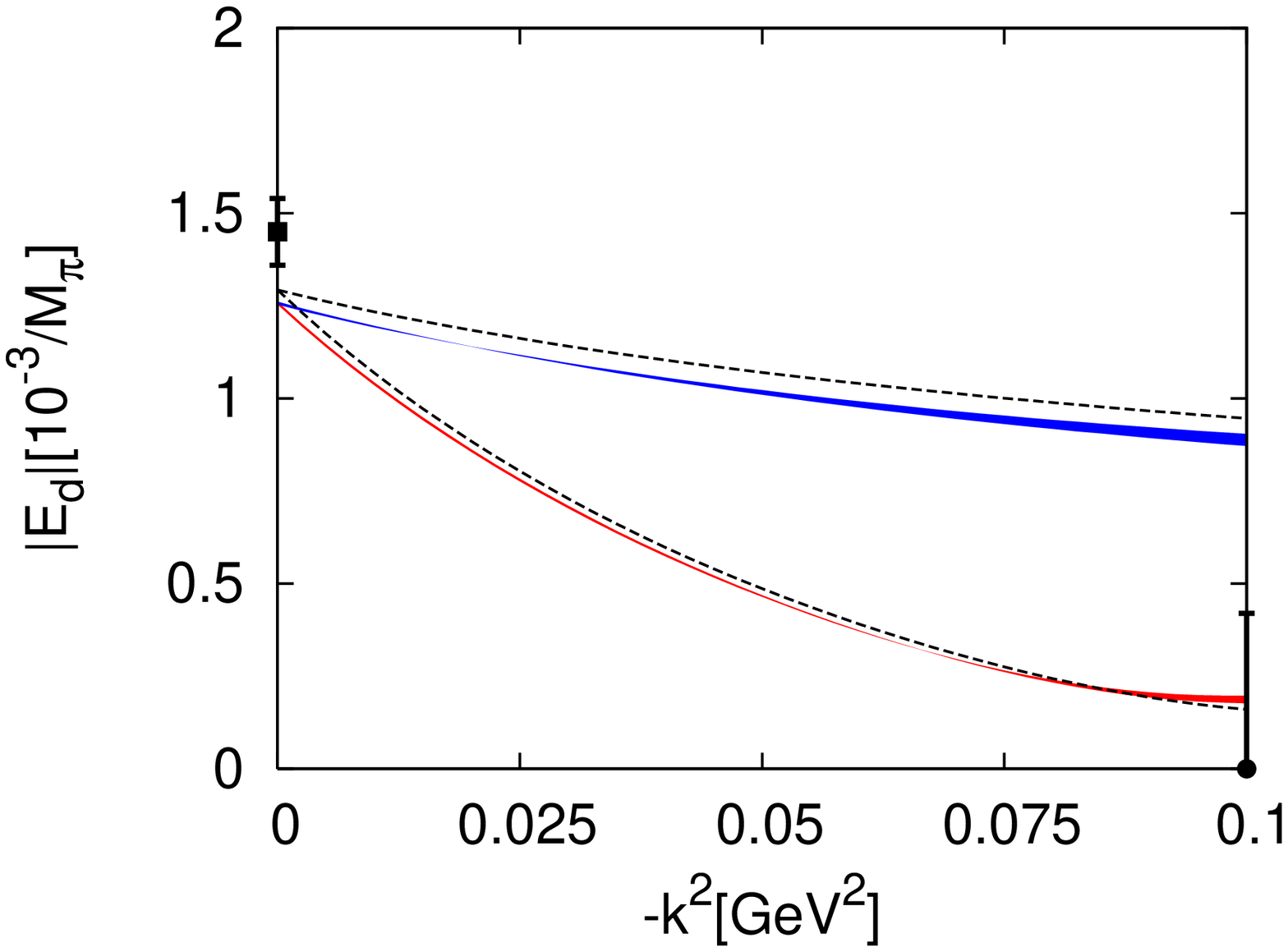}
    \includegraphics*[angle=270,width=0.45\textwidth]{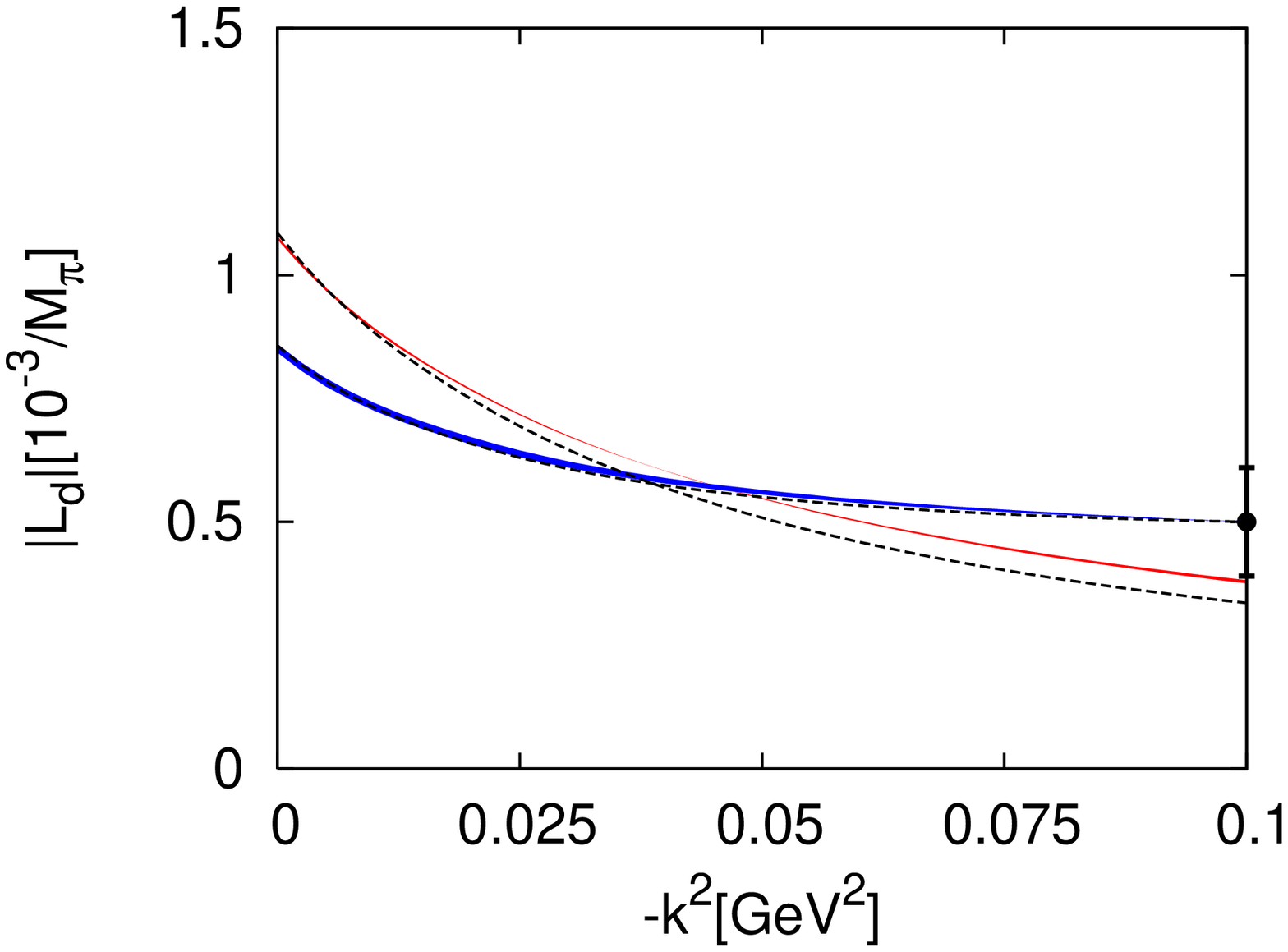}
   \vspace{0.7cm}
  \end{center} 
   \centerline{\parbox{12cm}{\caption{\label{fig:ELd} 
   Threshold multipoles $|E_d|$ (left panel) and $|L_d|$ 
   (right panel) as a function of the 
   photon virtuality in comparison to the photon point data 
   from SAL \protect\cite{SALd} and the electroproduction data from 
   MAMI  \protect\cite{Ewald}. The sign of the experimental result 
   for $L_d$ is taken to agree with the theoretical prediction. 
   $E_d$: The blue (upper)/red  (lower) band refers to fit~1,2, in order
   (for $L_d$, the upper/lower refers to the value at $Q^2=0.1\,$GeV$^2$).
   The dashed lines are the corresponding results from \protect\cite{KBM2}.
  }}} 
  \vspace{0.3cm}
\end{figure*}

\begin{figure*}[htbp] 
   \vspace{0.5cm} 
   \centerline{\epsfig{file=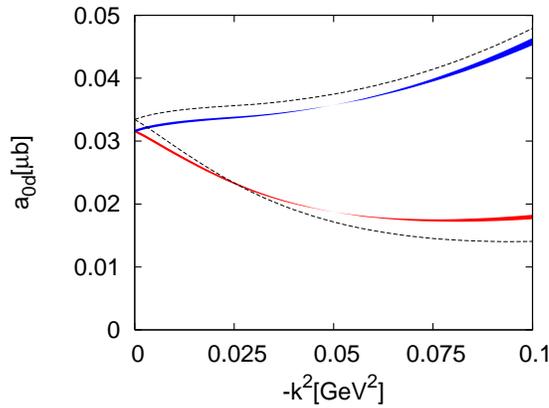,height=3in,angle=270}} 
   \vspace{0.3cm} 
   \centerline{\parbox{10cm}{\caption{\label{fig:a0d} 
   The S-wave  cross section $a_{0d}$  as a function of the 
   photon virtuality. Solid bands (dashed lines): Range obtained
   with the fourth (third) order three-body corrections.
   The dashed lines are the corresponding results from \protect\cite{KBM2}.
  }}} 
\end{figure*}

The corresponding total 
cross sections as a function of the excess energy $\Delta W$ and of the photon 
polarization $\varepsilon$ are shown in Figs.~\ref{fig:tot1}  
for fit~1 and in Figs.~\ref{fig:tot2}  for fit~2 
and the NNLO wave function with $\Lambda$ ranging from 450 to
650 MeV, respectively. 
We notice that for fit~1 with increasing excess energy and, in particular, with 
increasing photon polarization the data are systematically below the chiral  
prediction. Due to the fitting procedure, the slopes of the various curves 
for the Rosenbluth separation shown in the right panel of Fig.~\ref{fig:tot1} 
are of course  correct.  These results are very similar to the ones 
obtained in \cite{KBM2} for fit~1. We note that the predictions for fit 
procedure 2 are visibly
improved compared to the third order calculation. This is further seen
by looking at the  transverse and the longitudinal threshold S-wave multipole 
shown in Fig.~\ref{fig:ELd}. The prediction for $|L_d|$ for fit~2 is now
within the error bar of the MAMI result (which was not the case in \cite{KBM2}).
We note that the real part of the predicted longitudinal amplitude is
negative, consistent with our findings in \cite{KBM2}.  
The resulting S-wave cross section $a_{0d}$ is shown
in Fig.~\ref{fig:a0d}. The theoretical uncertainty is a bit smaller than in
case of the third order tb calculation \cite{KBM2}.

\subsection{Results with fitted four-nucleon LECs}

As argued before, we will also perform fits leaving the values of the two
four-nucleon LECs as free parameters. Since as we noted before these operators
contribute very little to the S-waves, and not at all at threshold,
the fits of type 1 are performed in
two stages. First, $a_5^n$ is adjusted to give the proper value of $|L_d|$ at
$k^2 = -0.1\,$GeV$^2$. Then, the two four-nucleon LECs are fitted to the
total cross sections. The fits of type~2 are performed as before, all parameters
are fixed on the total cross sections. Having said this, we collect in 
table~\ref{tab:CLECs} the values of the various LECs for the two types of
fits. We note that the four--nucleon LECs come out much larger than from
the investigation of nucleon-nucleon scattering, cf. table~\ref{tab:Cvals}.
However, as we discussed earlier, leaving these LECs as free parameters
effectively subsumes some effects from the single nucleon P-wave
contributions, so that these fits should be considered indicative only.
\renewcommand{\arraystretch}{1.3} 
\begin{table}[htb] 
\begin{center} 
\begin{tabular}{|c||c|c|} 
\hline 
LECs   &  $\Lambda =450$~MeV  &  $\Lambda =650$~MeV  \\ 
\hline 
$a_3^n$ [GeV$^{-4}$]              &\phantom{$-$}4.500 &\phantom{$-$}4.751 \\
$a_4^n$ [GeV$^{-4}$]              & $-$8.549 & $-$8.573 \\
$C_{1S0}$ [$10^{-4}\,$GeV$^{-2}$] & $-$0.589 & $-$0.664 \\  
$C_{3S1}$ [$10^{-4}\,$GeV$^{-2}$] & $-$1.168 & $-$1.105 \\  
\hline  
$a_5^n$ [GeV$^{-5}$] & $-$38.71 & $-$37.05 \\  
$C_{1S0}$ [$10^{-4}\,$GeV$^{-2}$] & $-$0.325 & $-$0.387 \\  
$C_{3S1}$ [$10^{-4}\,$GeV$^{-2}$] & $-$0.762 & $-$0.890 \\  
\hline
\end{tabular}  
\vspace{0.1cm}
\caption{Values of the fitted  LECs for the NNLO wave functions 
from \protect\cite{EGMZ}.
The values for $a_{3,4}^n$ refer to the fits~2 with the
corresponding values for $C_{1S0}$ and $C_{3S1}$, whereas 
the corresponding $a_5^n$ belongs to the respective fits~1,
again with the corresponding four-nucleon LECs. 
\label{tab:CLECs}} 
\end{center} 
\end{table} 

\begin{figure*}[htbp] 
\begin{center}
   \vspace{0.5cm} 
    \includegraphics*[angle=270,width=0.45\textwidth]{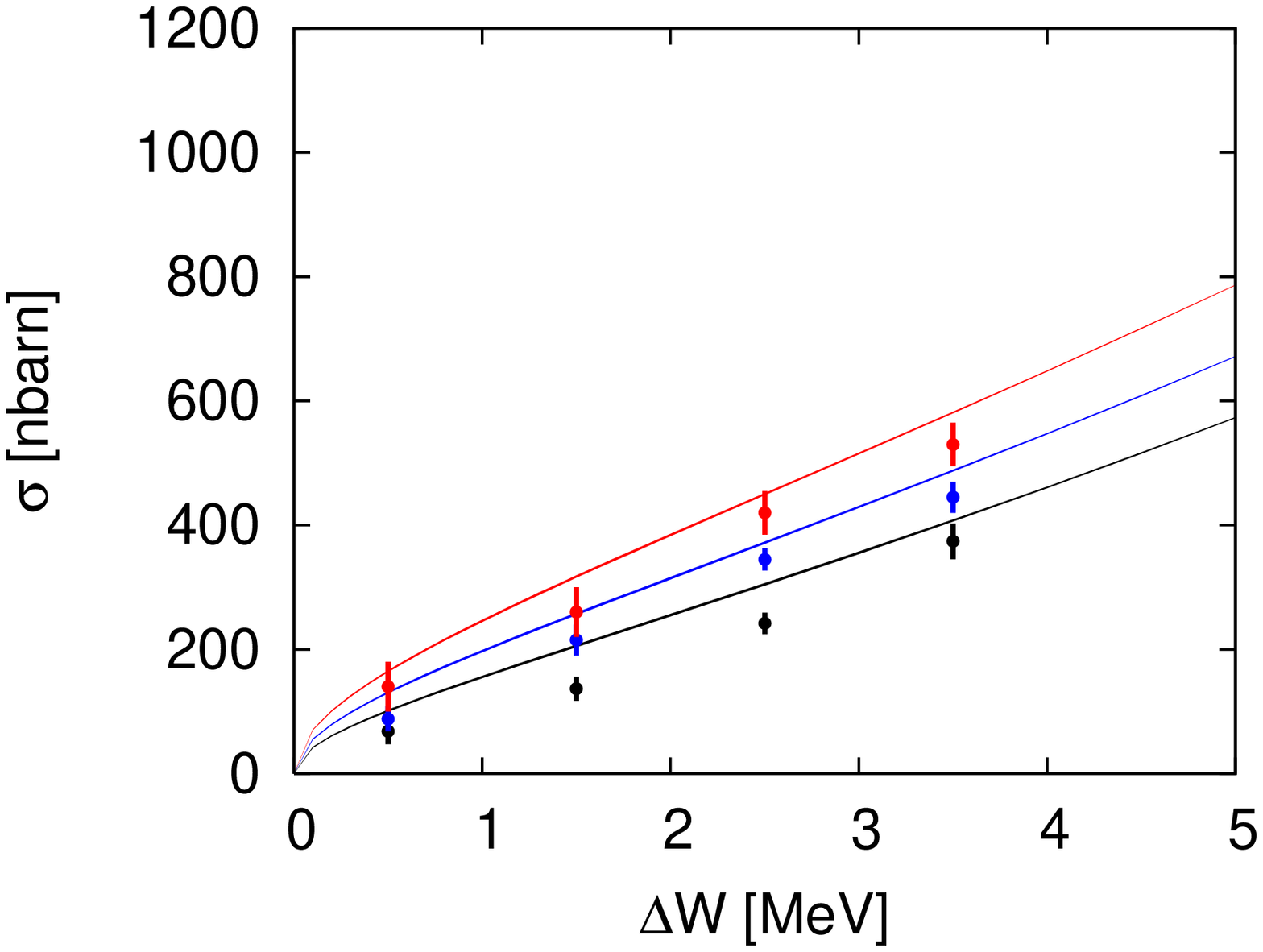}
    \includegraphics*[angle=270,width=0.45\textwidth]{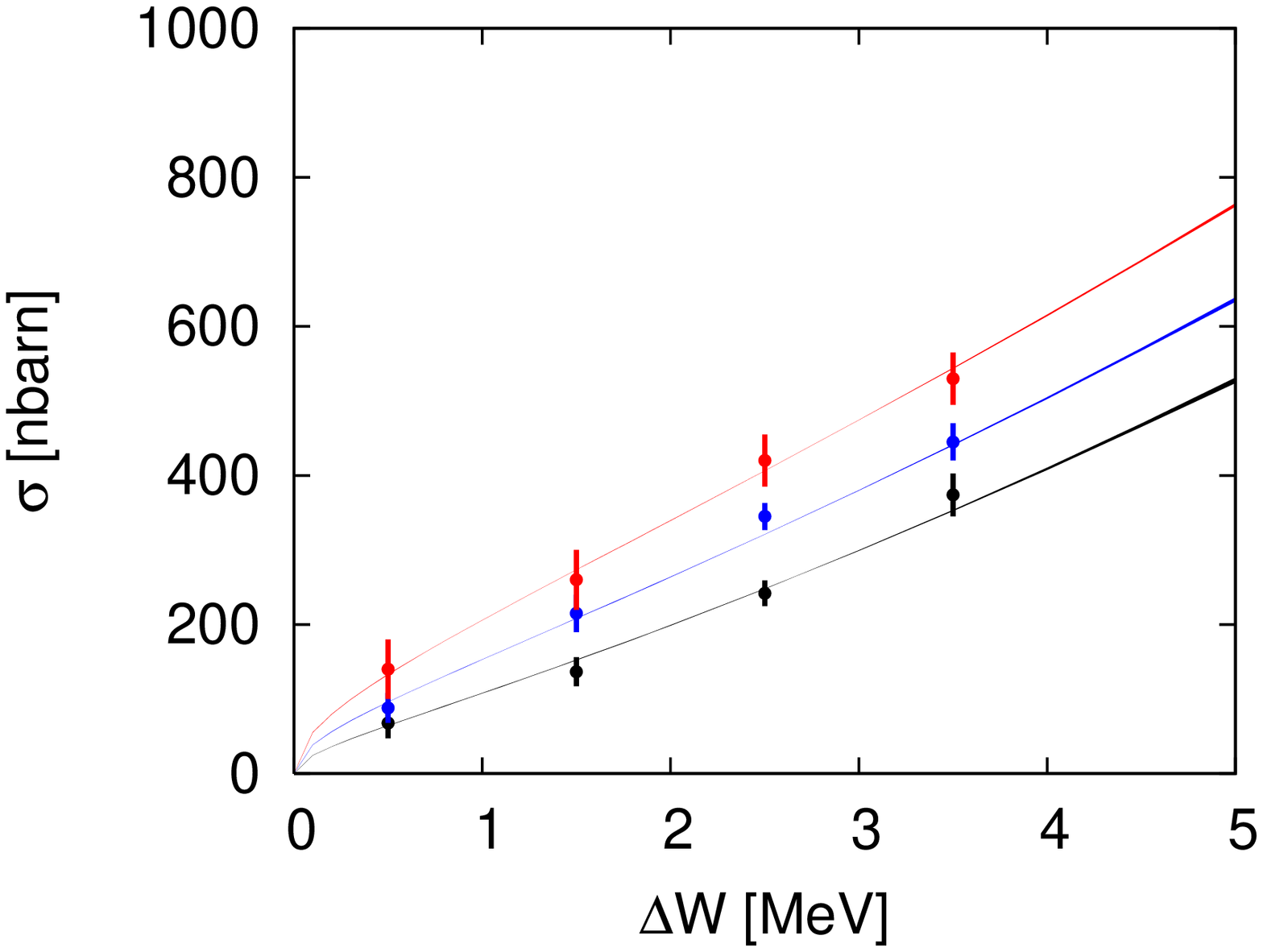}
   \vspace{0.7cm}
 \end{center} 
   \centerline{\parbox{12cm}{\caption{\label{fig:totC} 
   Left panel: Total cross section as a function of $\Delta W$ for three 
   different values of the photon polarization in comparison 
   to the MAMI data \protect\cite{Ewald} for fit~1 and the 
   NNLO wave functions. 
   The upper/middle/lower band corresponds to the largest/medium/smallest 
   value of $\varepsilon$. 
   Right panel: Same for fit~2.
  }}} 
\end{figure*}

\begin{figure*}[htbp] 
   \vspace{0.5cm} 
  \begin{center}
    \includegraphics*[angle=270,width=0.45\textwidth]{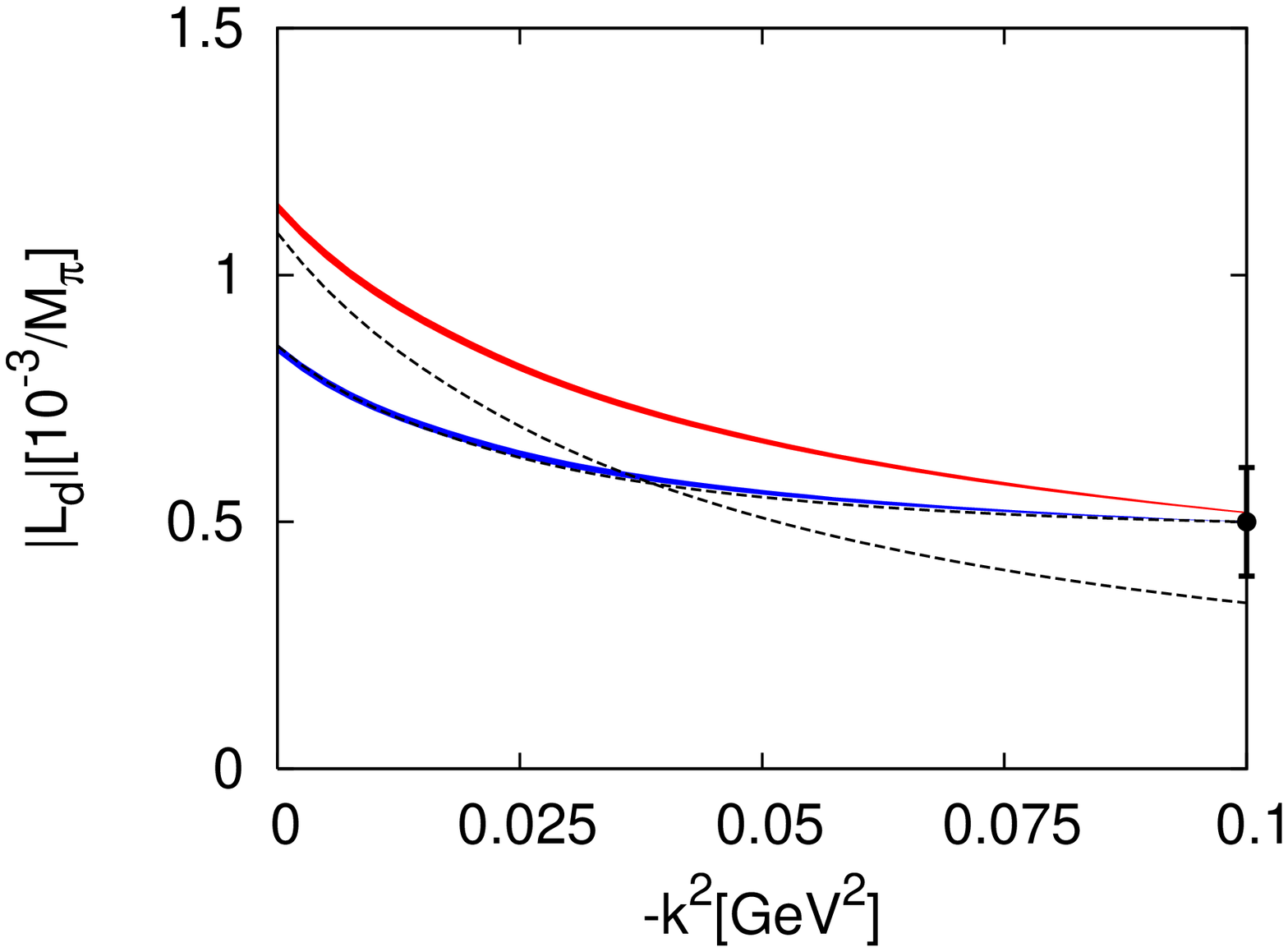}
    \includegraphics*[angle=270,width=0.45\textwidth]{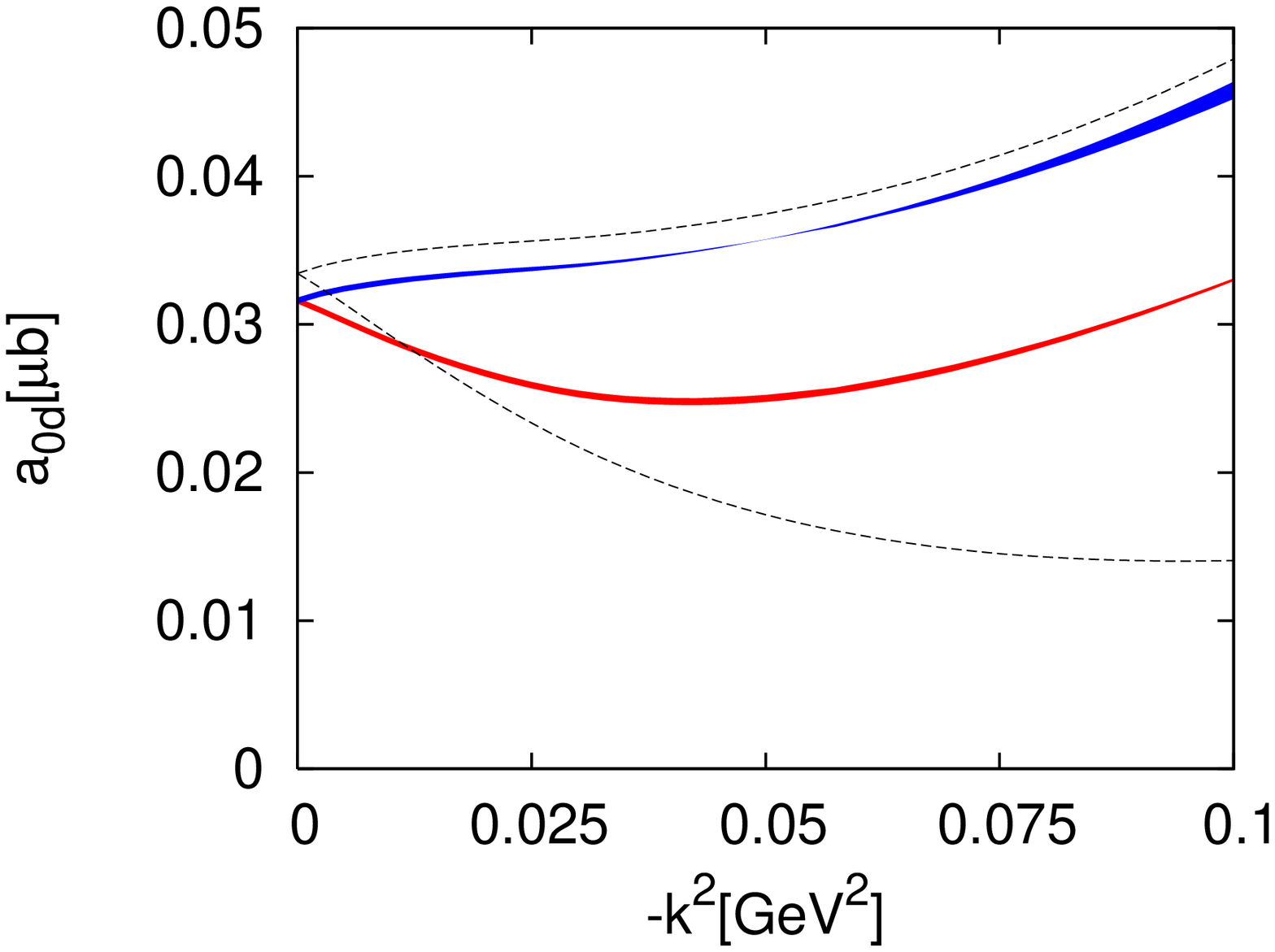}
   \vspace{0.7cm}
  \end{center} 
   \centerline{\parbox{12cm}{\caption{\label{fig:ELdC} 
   Fits with varying four-nucleon LECs.
   Left panel: Threshold multipole $|L_d|$ as a function of the 
   photon virtuality in comparison to the electroproduction data from 
   MAMI  \protect\cite{Ewald}. The sign of the experimental result 
   for $L_d$ is taken to agree with the theoretical prediction. 
   The blue (upper)/red  (lower) band refers to fit~1,2, in order.
   Right panel:  The S-wave  cross section $a_{0d}$  as a function of the 
   photon virtuality. The solid bands give the range  obtained
   with the fourth order three-body corrections, dashed lines show the
   third order tb results of \protect\cite{KBM2}.
  }}} 
  \vspace{0.3cm}
\end{figure*}

The resulting total cross sections are shown in Fig.~\ref{fig:totC}.
We observe a clear improvement for fit~1 and a very good description
for fit~2. We remark, however, that the corresponding differential 
cross sections for $\Delta W \ge 1.5\,$MeV are more symmetric around
$\cos(\theta)= 0$ (bell-shaped) than the ones shown in the preceding
section, whereas the data indicate a peaking into the backward direction. 
This points towards an insufficient accuracy in the description of
some of  the P-wave multipoles. The resulting longitudinal 
deuteron S--wave multipole is shown in Fig.~\ref{fig:ELdC}. While
$|E_d|$ is almost unaffected,
$|L_d (k^2 = -0.1\,{\rm GeV}^2)|$ for fit~2 agrees with the empirical
value. Consequently, the band for the S--wave cross section is much
narrower than before, cf. the right panel of Fig.~\ref{fig:ELdC}. 
This is because the cross sections are much sensitive to the longitudinal 
S-wave amplitude due to the kinematical enhancement proportional to the
longitudinal polarization $\varepsilon_L \simeq 9$, see \cite{Ewald}.

\section{Summary and outlook} 
\label{sec:sum}
 
In this manuscript, we have considered neutral pion electroproduction
off deuterium in the framework of heavy baryon chiral perturbation theory,
extending and improving upon the work presented in \cite{KBM2}. The salient
results of this study are:
\begin{itemize}
\item[i)]We have calculated the fourth order three-body corrections. These
consist of 51 one-pion exchange diagrams with exactly one insertion from the
dimension two pion--nucleon Lagrangian and 8 diagrams with one insertion from
the lowest order four--nucleon interaction Lagrangian. In principle, all parameters
are fixed from earlier studies of pion--nucleon and nucleon--nucleon scattering.
The deuteron wave functions are taken consistently from the recent chiral
EFT study of \cite{EGMZ}.
\item[ii)]As in the earlier work \cite{KBM2}, in which the three-body corrections
were only considered at third order, we have performed two types of fits to the
MAMI data \cite{Ewald}. The results are similar to the ones found there, although
we observe a moderate improvement in the theoretical uncertainty. This is most
pronounced for the longitudinal S-wave multipole $|L_d|$ in fit~2. We have also
considered the dependence on the deuteron wave functions related to the cut--off
in the Lippmann-Schwinger equation. It turned out that this dependence
is very weak. Therefore, the process is dominated by long--range pion physics
and thus sensitive to the elementary $\pi^0 n$ amplitude.
\item[iii)]We have also performed fits where we have left the four--nucleon
LECs as free parameters, based on the argument that in that way one can effectively
subsume new P-wave LECs from the single scattering contribution. This leads to 
a visibly improved description of the total cross sections and the longitudinal S--wave
multipole. However, the differential cross sections come out bell-shaped, in contrast
to the experimental findings \cite{Ewald}.
\end{itemize}
 
In conclusion, we have further sharpened the theoretical framework to analyse
neutral pion production off deuterium in a model--independent framework. It remains
to be seen whether a further improved description of the data will be possible
when the fourth order pion--nucleon P--wave multipoles at fourth order become
available.

\section*{Acknowledgements}
We thank Evgeny Epelbaum for useful comments and supplying us
with the EFT wave functions.

\appendix 
\def\theequation{\Alph{section}.\arabic{equation}} 
\setcounter{equation}{0} 
\section{Fourier transformations}
\label{app:fourier} 
The diagrams are evaluated in momentum space. To calculate the pertinent
matrix elements, we perform Fourier integrations, see e.g. \cite{BBLMvK}.
The typical structures to be evaluated take the form
\beqa
&&p_{\nu 1}' \ldots p_{\nu m}'\, \Phi^\star \left( \vec{p}\,'\right)
\no\\ &=& {1 \over
(2\pi)^{3/2}} \int d^3r_1 \,{\rm e}^{-i \vec{p}\,'\cdot \vec{r}_1}\,
(-i\partial_{\nu 1})\ldots  (-i\partial_{\nu m}) \phi\left(  \vec{r}_1 \right)
\no\\ 
&&p_{\sigma 1} \ldots p_{\sigma k} \, \Phi \left( \vec{p}\,\right)
\no\\ &=& {1 \over
(2\pi)^{3/2}} \int d^3r_2 \,{\rm e}^{i \vec{p}\cdot \vec{r}_2}\,
(i\partial_{\sigma 1})\ldots  (i\partial_{\sigma k}) \phi\left(  \vec{r}_2 \right)~,
\eeqa
where $p \, (p')$ are initial (final) state relative nucleon momenta, 
$\phi$ is the
coordinate space deuteron wave function and similarly $\Phi$ denotes the
momentum space wave function. As stated, we have to consider 5 classes of
diagrams,
\begin{itemize}
\item[1)]Diagrams of type a) with $q_0'= q_0 + {\mathcal O}(1/m)$,
\item[2)]Diagrams of type a) with $q_0'=  {\mathcal O}(1/m)$,
\item[3)]Diagrams of type b),
\item[4)]Time-ordered diagrams of type c),
\item[5)]The NN-diagrams of type d).
\end{itemize}
In the following, $\Theta$ symbolizes an arbitrary spin-operator.

\medskip
Consider first the diagrams 1). The momentum of the exchanged pion
is 
\be\label{A2}
\vec{q}\,' = \vec{p} - \vec{p}\,' + { \vec{k}\over 2}  + { \vec{q}\over 2}~.
\ee
From the momentum space expression for this class of diagrams, one
deduces the following  pertinent Fourier-transform 
\be
{1 \over q'\cdot q' + \omega_c^2 -q_0^2 - i\epsilon} 
= {1 \over4\pi} \int d^3r \, {\rm e}^{-i \vec{q}\,'\cdot \vec{r}}\,
{{\rm e}^{-\delta \, r}\over r}~,
\ee
with
\be
\delta = \omega_c \,\sqrt{1 - {q_0^2\over \omega_c^2}}~.
\ee
Here, $\omega_c$ is the physical value of the $nn\pi^+$ and
the $pp\pi^-$ threshold, respectively. So then we have to
evaluate the following integrals
\beqa
&&\int d^3p \, d^3p' \, p_{\nu 1}' \ldots p_{\nu m}'\, 
\Phi^\star \left( \vec{p}\,'\right)
{1 \over q'\cdot q' + \omega_c^2 -q_0^2 - i\epsilon} \Theta\no\\
&&\times  
p_{\sigma 1} \ldots p_{\sigma k} \, \Phi \left( \vec{p}\,\right)
= 2\pi^2 \int d^3r (-i\partial_{\nu 1})\ldots  (-i\partial_{\nu m}) 
\phi\left(  \vec{r} \right)\no\\
&&\times \Theta {{\rm e}^{-\delta \, r}\over r} 
i\partial_{\sigma 1}\ldots  i\partial_{\sigma k}\, \phi\left(  \vec{r} \right)
\, {\rm e}^{-i \left({\vec{k}\over 2}+{\vec{q}\over 2}\right)\cdot \vec{r}}~.
\eeqa

\medskip
Next, we evaluate the diagrams 2). The momentum of the exchanged pion now is 
\be\label{moms}
\vec{q}\,' = \vec{p} - \vec{p}\,' + { \vec{k}\over 2}  - { \vec{q}\over 2}~.
\ee
and the pertinent Fourier integral is
\be
{1 \over q'\cdot q' + M_{\pi^+}^2  - i\epsilon} 
= {1 \over4\pi} \int d^3r \, {\rm e}^{-i \vec{q}\,'\cdot \vec{r}}\,
{{\rm e}^{- M_{\pi^+} \, r}\over r}~,
\ee
leading to the structures
\beqa
&&\int d^3p \, d^3p' \, p_{\nu 1}' \ldots p_{\nu m}'\, 
\Phi^\star \left( \vec{p}\,'\right)
{1 \over q'\cdot q' + M_{\pi^+}^2  - i\epsilon}\, \Theta\no\\
&&\times  
p_{\sigma 1} \ldots p_{\sigma k} \, \Phi \left( \vec{p}\,\right)
= 2\pi^2 \int d^3r (-i\partial_{\nu 1})\ldots  (-i\partial_{\nu m}) 
\phi\left(  \vec{r} \right)\no\\
&&\times \Theta {{\rm e}^{-M_{\pi^+} \, r}\over r} 
i\partial_{\sigma 1}\ldots  i\partial_{\sigma k}\, \phi\left(  \vec{r} \right)
\, {\rm e}^{-i \left({\vec{k}\over 2}-{\vec{q}\over 2}\right)\cdot \vec{r}}~.
\eeqa

\medskip
We now consider the diagrams 3). Here, the photon couples to the pion in
flight, thus we have two propagators. The momentum of the exchanged pion
before the photon absorption is
\be\label{A9}
\vec{q}\,' = \vec{p} - \vec{p}\,' + { \vec{k}\over 2}  + { \vec{q}\over 2}~,
\ee
and after the photon absorption this momentum is changed to
\be
\vec{q}\,'' = \vec{q}\,' - \vec{k}~.
\ee
We have two types of Fourier-transforms, these are:
\beqa
{1 \over (q''\cdot q'' + \omega_c^2 - i\epsilon) (q'\cdot q' + \omega_c^2 -q_0^2
 - i\epsilon) } \no\\
= \int_0^1 dz {1 \over \left( {\vec l}\,^2+{m'}^2  - i\epsilon
\right)^2}~,
\eeqa
with 
\beqa
\vec{l} &=& \vec{q}\,' - z \, \vec{k}~, \no\\
{m'}^2  &=& \omega_c^2 \, z + (1-z)\, \left( \delta^2 + \vec{k}\,^2 \, z
\right)~,
\eeqa
and
\beqa 
{q_{\mu 1}' \, q_{\mu 2}' \over (q''\cdot q'' + \omega_c^2 - i\epsilon) 
   (q'\cdot q' + \omega_c^2 -q_0^2
 - i\epsilon) } \qquad && \no\\
= \int_0^1 dz {1 \over 8 \pi m'} \,\int d^3r \,  {\rm e}^{-i \vec{l} \cdot \vec{r}}
 \qquad\qquad\qquad\qquad\qquad &&
\no \\
\times \, (-i \partial_{\mu 1} + z  k_{\mu 1}) (-i \partial_{\mu 2} + z  k_{\mu 2})\, 
 {\rm e}^{- m' \, {r}}\,.  \qquad &&
\eeqa
Note that this function is still square integrable. Higher order derivatives, however,
can not be treated in this fashion. The corresponding integrals are thus
\beqa
&&\int d^3p \, d^3p' \, p_{\nu 1}' \ldots p_{\nu m}'\, 
\Phi^\star \left( \vec{p}\,'\right)\no\\
&&\times {q_{\mu 1}' \, q_{\mu 2}' \over 
(q''\cdot q'' + \omega_c^2 - i\epsilon) (q'\cdot q' + \omega_c^2 -q_0^2
 - i\epsilon) }  \no\\
&&\times \Theta \, 
p_{\sigma 1} \ldots p_{\sigma k} \, \Phi \left( \vec{p}\,\right)
= \int_0^1 dz \,{\pi^2\over m'} \int d^3r \, \no\\
&& \times (-i\partial_{\nu 1})\ldots  (-i\partial_{\nu m}) 
\phi\left(  \vec{r} \right) \,\Theta \, (-i \partial_{\mu 1} + z  k_{\mu 1})\no\\
&&\times (-i \partial_{\mu 2} + z  k_{\mu 2}) \, {\rm e}^{-m' \, r}\, 
i\partial_{\sigma 1}\ldots  i\partial_{\sigma k}\, \phi\left(  \vec{r} \right)
\no\\
&&\times {\rm e}^{-i \left({\vec{q}\over 2}- \left(z - \frac{1}{2}\right)\vec{k}\right)
\cdot \vec{r}}~.
\eeqa

\medskip\noindent
We now discuss the diagrams 4), which have the momentum transfer given either 
by Eq.~(\ref{A2}) or by Eq.~(\ref{moms}). Due to the
cancellation with the corresponding Okubo-type diagrams, no new structures
as compared to the other type of diagrams appears (for a more detailed
discussion on this point see \cite{KBMokubo})

\medskip\noindent
Finally, we turn to the NN  diagrams 5). Since these only involve contact interactions,
we have to consider simple structures of the form
\beqa
&&\int d^3p \, d^3p' \, p_{\nu 1}' \ldots p_{\nu m}'\, 
\Phi^\star \left( \vec{p}\,'\right)
\, \Theta \, 
p_{\sigma 1} \ldots p_{\sigma k} \, \Phi \left( \vec{p}\,\right)\no\\
&=& (2\pi)^3\, (-i\partial_{\nu 1})\ldots  (-i\partial_{\nu m}) 
\phi\left(  \vec{r}_1 \right) \bigl|_{{r_1}=0}\,\Theta \, \no\\
&\times& (i \partial_{\sigma 1})\ldots
(i \partial_{\sigma k}) \,  \phi\left(  \vec{r}_2 \right)\bigl|_{{r_2}=0}
\eeqa

\medskip
We end this appendix with a note on how we treat the contributions arising
from the exchange of nucleon 1 with nucleon 2 when  calculating the pertinent
amplitude, $A = A(1,2) + (1\leftrightarrow 2)$. The corresponding momenta of
the nucleons in terms of the center-of-mass and the photon and pion momenta
are
\beqa
\vec{p}_1  &=& \vec{p} -  {\vec{k}\over 2}~, \quad
\vec{p}_1'  = \vec{p}' -  {\vec{q}\over 2}~, \no\\
\vec{p}_2  &=& -\vec{p} -  {\vec{k}\over 2}~, \quad
\vec{p}_2'  = -\vec{p}' -  {\vec{q}\over 2}~. 
\eeqa
Under a parity transformation $ \vec{p} \to -\vec{p}$ and
$ \vec{p}' \to -\vec{p}'$, the nucleon momenta behave as
$\vec{p}_1 \to \vec{p}_2~$, $\vec{p}_2 \to \vec{p}_1~$, 
$\vec{p}_1' \to \vec{p}_2'~$, $\vec{p}_2' \to \vec{p}_1'~$, which gives
\beqa
&&\int d^3 p  d^3p'  \Phi^\star \left( \vec{p}'\right)
f\left( \vec\sigma_1,\vec\sigma_2,\vec{p}_1,\vec{p}_2,\vec{p}_1',\vec{p}_2'
\right) \Phi\left( \vec{p}\right) \no\\
&=&\int d^3 p  d^3p' \Phi^\star \left( \vec{p}'\right)
f\left( \vec\sigma_1,\vec\sigma_2,\vec{p}_2,\vec{p}_1,\vec{p}_2',\vec{p}_1'
\right) \Phi\left( \vec{p}\right) .\no\\&&
\eeqa
Thus, matrix elements are invariant under operations of the form
\beqa
&&
f\left( \vec\sigma_1,\vec\sigma_2,\vec{p}_1,\vec{p}_2,\vec{p}_1',\vec{p}_2'
\right)\to \no\\
&&\quad
f\left( \vec\sigma_1,\vec\sigma_2,\vec{p}_2,\vec{p}_1,\vec{p}_2',\vec{p}_1'
\right)\, .\no
\eeqa
Consequently, the operation $(1 \leftrightarrow 2)$ need only be applied to
the spin and isospin operators and leave all other quantities unchanged. 
Furthermore, because of isospin symmetry we only have the operators 
$\tau_1 \cdot \tau_2 - \tau_1^z  \tau_2^z$ and $ \tau_1^z  \tau_2^z$, which
are symmetric under the interchange  $(1 \leftrightarrow 2)$. Therefore, we
finally only need to change the spin indices when calculating the
contributions from the exchange diagrams.

\setcounter{equation}{0} 
\section{Angular integrations}
\label{app:angular} 
Here, we briefly discuss how to efficiently perform
the angular integrations of the Fourier
integrals discussed in App.~\ref{app:fourier}. We define the following matrices constructed
from the momentum space deuteron wave functions,
\beqa\label{wfmatrix}
u_{n,m}(r) &=& {1\over r^{n+m}} \sqrt{\frac{2}{\pi}} \int_0^\infty dq \, q^2 \, u(q)\,
(q r)^n \, j_m (qr)~,\no \\
w_{n,m}(r) &=& -{1\over r^{n+m}} \sqrt{\frac{2}{\pi}} \int_0^\infty dq \, q^2 \, w(q)\,
(q r)^n \, j_m (qr)~, \no\\ &&
\eeqa
where $u(q)$ and $w(q)$ are the momentum-space S- and D-wave components of the deuteron
wave function, and the $j_n (qr)$ are spherical Bessel functions. The usual 
coordinate space expressions can be obtained using matrix elements
$u_{0,0}$ and $w_{0,2}$, i.e. 
\be 
u_{0,0} = {u(r) \over r} ~,  \,  w_{0,2} = {w(r) \over r^3} ~.
\ee
This notations is particularly useful since we have to deal with derivatives of the
wave functions. Using the properties of the spherical Bessel functions, one obtains the
following recursion relation
\be
r^2 u_{n+1,m+2}(r) -(2m+3)u_{n,m+1}(r) + u_{n+1,m}(r) = 0~.
\ee
This allows one to express the derivatives of $u_{0,0}$ in the following way
\beqa
u_{0,0}' (r)   &=& -r \, u_{1,1} (r)~, \no\\
u_{0,0}'' (r)  &=& - u_{1,1} (r) + r^2 \,u_{2,2} (r) ~, \no\\
u_{0,0}''' (r) &=& 3r \, u_{2,2} (r) - r^3 \,u_{3,3} (r) ~.
\eeqa
Similarly, the derivatives of $w_{0,2}$ can be expressed as
\beqa
w_{0,2}' (r)   &=& -{1\over r} \, \left( 5w_{0,2} (r) - w_{1,1} (r)\right)~, \no\\
w_{0,2}'' (r)  &=& {1\over r^2} \, \left( 30w_{0,2} (r) - 6w_{1,1} (r)
                                          -r^2 w_{22}(r)\right)\no\\ 
w_{0,2}''' (r)  &=& {1\over r^3} \, \left( -210w_{0,2} (r) + 42w_{1,1} (r)\right.\no\\
               && \qquad\qquad  \left. +6r^2w_{2,2} (r) + r^4 w_{3,3} (r) \right)~.
\eeqa
Note that one only has to calculate diagonal matrix elements.


\newpage

\vspace{1.2cm}
 
 
\vspace{2cm}
 
\vspace{1.2cm}

\end{document}